\documentclass[journal]{IEEEtran}
\usepackage{amsmath, amsfonts, amssymb}
\usepackage{array}
\usepackage[caption=false,font=normalsize,labelfont=sf,textfont=sf]{subfig}
\usepackage{textcomp}
\usepackage{stfloats}
\usepackage{url}
\usepackage{verbatim}
\usepackage{graphicx}
\def\BibTeX{{\mathrm B\kern-.05em{\sc i\kern-.025em b}\kern-.08em
    T\kern-.1667em\lower.7ex\hbox{E}\kern-.125emX}}
\usepackage{balance}

\usepackage{bm}         
\usepackage{booktabs}   
\usepackage{tabularx}   
\usepackage{cite}       
\usepackage{algorithm}  
\usepackage{algpseudocode}
\usepackage{setspace}   
\usepackage{mathrsfs}   
\usepackage{cuted}      
\usepackage{booktabs}
\usepackage{multirow}
\usepackage{graphicx}
\usepackage{threeparttable}

\begin{document}

\title{SA-GLRT-D: Sidelobe-Aware GLRT-Based Detector for FMCW Radar}
\author{Jiachen Zhu, Fangjiong Chen,~\IEEEmembership{Member,~IEEE}, Jie Wu, and Ming Xia
\thanks{Manuscript received XXXXX 00, 0000; revised XXXXX 00, 0000; accepted XXXXX 00, 0000. 
Date of publication XXXXX 00, 0000; date of current version XXXXX 00, 0000. 
This work was supported in part by the National Natural Science Foundation of China under Grant 62271208, 
in part by the Basic and Applied Basic Research Foundation of Guangdong Province under Grant 2024A1515011107, 
in part by the Guangxi Key Research and Development Program under Grant AB25069510, 
and in part by the Key Technologies Research and Development Program of Guangzhou Municipality under Grant 2023B03J1308.
\emph{(Corresponding author: Fangjiong Chen.)}}
\thanks{Jiachen Zhu and Fangjiong Chen are with the School of Electronic and Information Engineering, 
South China University of Technology, Guangzhou 510641, China, 
and also with the Guangdong Provincial Key Laboratory of Short-Range Wireless Detection and Communication, 
South China University of Technology, Guangzhou 510641, China (e-mail: eezhujc@mail.scut.edu.cn; eefjchen@scut.edu.cn).}
\thanks{Jie Wu is with Southern Marine Science and Engineering Guangdong Laboratory (Guangzhou), 
Guangzhou 511458, China (e-mail: wu\_jie@gmlab.ac.cn).}
\thanks{Ming Xia is with the School of Electronics and Information Engineering, 
Beihang University, Beijing 100083, China (e-mail: xiaming@buaa.edu.cn).}
}

\markboth{IEEE TRANSACTIONS ON SIGNAL PROCESSING, ~Vol.~xx, No.~x, May~2026}
{SA-GLRT-D: Sidelobe-Aware GLRT-Based Detector for FMCW Radar}
\maketitle
\begin{abstract}
This paper investigates adaptive detection for frequency-modulated continuous-wave (FMCW) radar in multi-target complex scenarios, 
where reference-cell contamination, spectral leakage from the cell under test (CUT), and the resulting sidelobe interference severely degrade detection performance. 
A sidelobe-aware detector based on the generalized likelihood ratio test (GLRT), termed SA-GLRT-D, is proposed. Unlike conventional filtering-then-constant false alarm rate (CFAR) schemes,
the proposed method explicitly accounts for the non-ideal frequency-domain energy spreading of the CUT induced by the finite-window effect. 
Accordingly, conventional single-cell target detection is extended to local-region detection, and decision statistic construction, CUT-related parameter estimation,
and sidelobe interference suppression are jointly accomplished within a unified hypothesis-testing framework. Furthermore, to reduce computational complexity, a fast implementation of SA-GLRT-D is developed,
together with a dedicated parameter initialization strategy to enhance the stability of numerical optimization. 
Simulation and real-measurement results demonstrate that the proposed method achieves high-accuracy target detection while maintaining a constant false alarm rate, with favorable computational efficiency.
\end{abstract}

\section{Introduction}
\IEEEPARstart{F}{requency}-modulated continuous-wave (FMCW) rada-r, owing to its high resolution, strong environmental adaptability, and good anti-interference capability 
\cite{1.Wei2023AreaBasedCFARTargetDetection,
2.Zhang2025mmRotation}, 
has been widely used as an important sensing modality in fields such as security surveillance 
\cite{3.Dai2024AdaptiveTrackingStrategyForThePositioningOfMillimeterWaveRadarSecurityRobots,
4.Sebestyen2016HumanActivityRecognitionAndMonitoringForElderlyPeople,
5.Longo2023AttackingAndDefendingTheMaritimeRadarSystem}, 
autonomous driving 
\cite{6.Yang2025RaLiBEVRadarAndLiDARBEVFusionLearningForAnchorBoxFreeObjectDetectionSystems,
7.Schramm2024BEVCarCameraRadarFusionForBEVMapAndObjectSegmentation,
8.Fu2024CRPlaceCameraRadarFusionWithBEVRepresentationForPlaceRecognition,
9.Ma2020JointRadarCommunicationStrategiesForAutonomousVehiclesCombiningTwoKeyAutomotiveTechnologies}, 
vital sign monitoring 
\cite{10.Paterniani2023RadarBasedMonitoringOfVitalSignsATutorialOverview,
11.Li2013AReviewOnRecentAdvancesInDopplerRadarSensorsForNoncontactHealthcareMonitoring}, 
and human activity recognition
\cite{12.Zhu2025mRadHPRSHumanPoseRecognitionSystemFromPointCloudsGeneratedThroughAMillimeterWaveRadar,
13.Zhu2024RobustCFARDetectorBasedOnCLEANForSidelobeSuppression,
14.Zhu2025ProbRadarM3FmmWaveRadarBasedHumanSkeletalPoseEstimationWithProbabilityMapGuidedMultiformatFeatureFusion,
15.Zhou2025HumanActivityRecognitionBasedOnTheMethodOfCFAR-SOPCUsingMillimeterWaveRadar}. 
However, with the increasing complexity of application scenarios, achieving accurate and robust target detection in multi-target environments has become a critical challenge that restricts further performance improvement of FMCW radar systems
\cite{16.Wang2024JointDetectionAndDelayDopplerEstimationAlgorithmsForMIMORadars,
18.Huang2024AdaptiveAlgorithmGenerating3DPointCloudsHumanBodyBased4DMillimeterWaveRadar,
17.Wang2025AnalysisOfMIMORadarDetectionAlgorithmsWithLocationCapabilities,
19.Wang2026BayesianInterferenceControlCFARDetectorInGammaDistributedBackgroundUsingDiscreteTruncatedGaussianFunction}.

Target detection in FMCW radar commonly follows the constant false alarm rate (CFAR) framework, 
which can be broadly categorized into non-adaptive filtering followed by CFAR, adaptive filtering followed by CFAR, and adaptive filtering-based detection 
\cite{20.Liu2022MultichannelAdaptiveSignalDetectionBasicTheoryAndLiteratureReview}. 
For the first category, classical mean-based detectors such as CA-CFAR are optimal in homogeneous exponentially distributed clutter 
but degrade when reference cells are contaminated by outliers 
\cite{21.Gandhi1994OptimalityOfTheCellAveragingCFARDetector}. 
Representative extensions, 
including CAGO-CFAR \cite{22.Hansen1973ConstantFalseAlarmRate}, 
CASO-CFAR \cite{23.Trunk1978RangeResolutionOfTargetsUsingAutomaticDetectors}, 
OS-CFAR \cite{26.Blake1988OSCFARTheoryForMultipleTargetsAndNonuniformClutter}, 
TM-CFAR \cite{27.Gandhi1988AnalysisOfCFARProcessorsInNonhomogeneousBackground}, 
ICCA-CFAR \cite{30.Cao2021CompressedSensingBasedMultitargetCFARDetectionAlgorithmForFMCWRadar}, 
and ICCASO-CFAR \cite{31.Ding2016RobustCFARDetectorBasedOnTruncatedStatisticsInMultipleTargetSituations}, 
improve robustness under clutter edges and multi-target interference by using split-window, 
ordered-statistic, trimming, or iterative-censoring strategies.

Although non-adaptive filtering followed by CFAR methods provide low-complexity robustness against non-homogeneous reference cells, 
their ability to suppress local interference around the cell under test (CUT) remains limited 
\cite{32.Wu2026DetectorDesignAndPerformanceAnalysisForAdaptiveCFARDetectionInNonzeroMeanEnvironment}. 
This limitation becomes critical when strong neighboring targets or window-induced spectral leakage contaminate the reference cells, 
making the homogeneity assumption difficult to recover \cite{13.Zhu2024RobustCFARDetectorBasedOnCLEANForSidelobeSuppression}. 
Adaptive filtering followed by CFAR methods therefore introduce a preprocessing stage before thresholding. 
Typical examples include truncated statistics (TS)-CFAR 
\cite{31.Ding2016RobustCFARDetectorBasedOnTruncatedStatisticsInMultipleTargetSituations}, 
which improves robustness to clutter edges and local outliers but may encounter unstable 
or nonexistent likelihood solutions under severe contamination 
\cite{33.Ding2016ASegmentationBasedCFARDetectionAlgorithmUsingTruncatedStatistics,
34.Zefreh2021RobustCFARDetectorBasedOnCensoredHarmonicAveragingInHeterogeneousClutter}. 
SS-CFAR, which exploits second-order difference statistics and CLEAN-based processing to suppress sidelobe-induced false alarms 
\cite{13.Zhu2024RobustCFARDetectorBasedOnCLEANForSidelobeSuppression,43.Hogbom1974ApertureSynthesisWithANonRegularDistributionOfInterferometerBaselines}.
CT-CFAR, which combines TS-based noise estimation with CLEAN-based sidelobe suppression for improved multi-target robustness 
\cite{35.Zhu2025CTCFARRobustCFARDetector}.
Despite their favorable detection performance, the filtering-then-CFAR paradigm, with or without adaptive preprocessing, 
separates interference suppression from CFAR decision-making and thus lacks joint optimization under a unified statistical model 
\cite{36.Qi2026CooperativeFiniteTransmitReceiveAntennaSelection}.
This limitation is critical in multi-target FMCW scenarios, 
where window-induced CUT energy spreading and the reliance of CLEAN-like methods on accurate echo estimation can degrade detection performance 
\cite{35.Zhu2025CTCFARRobustCFARDetector}. 
TS-GLRT-D addresses delay-Doppler-domain target energy leakage by extending single-cell detection to a local-region GLRT formulation for MIMO radar 
\cite{16.Wang2024JointDetectionAndDelayDopplerEstimationAlgorithmsForMIMORadars}. 
However, its statistic is relatively complicated, reference-cell outliers and interfering target components are not fully considered, 
and its threshold calibration is mainly conducted under ideal simulations, limiting its applicability to measured multi-target FMCW scenes.

A key unresolved issue is to jointly model interference suppression and target detection while maintaining practical threshold calibration 
and manageable hypothesis-testing complexity. 
This paper addresses this issue by proposing a sidelobe-aware GLRT detector, termed SA-GLRT-D, for FMCW radar. 
SA-GLRT-D extends single-cell detection to local-region detection through a three-stage GLRT-based framework, 
where the decision statistic and CUT-related parameters are derived from a unified hypothesis-testing model. 
CLEAN-inspired processing
\cite{43.Hogbom1974ApertureSynthesisWithANonRegularDistributionOfInterferometerBaselines} 
is then incorporated to suppress CUT-induced sidelobes, enabling joint interference suppression and target detection. 
A fast implementation, termed SA-GLRT-D\(_{\mathrm{F}}\), 
and parameter initialization schemes are further developed to reduce computational cost and improve numerical stability. 
To the best of the authors' knowledge, 
this is the first adaptive filtering-based FMCW radar detector that simultaneously accounts for reference-cell contamination 
and target-leakage-induced sidelobe interference.
The main contributions of this work are summarized as follows:

\begin{enumerate}
    \item An adaptive filtering-based CFAR detector is proposed for FMCW radar. To the best of the authors' knowledge, under the considered FMCW/NCA setting, 
    this is the first practical detector in the open literature to jointly address reference-cell contamination and CUT-induced sidelobe interference.

    \item A fast implementation of the proposed SA-GLRT-D, termed SA-GLRT-D\(_{\mathrm{F}}\), is further developed. Low-order analytical approximations and tailored parameter initialization are 
    introduced to improve computational efficiency and ensure numerical stability.
    
    \item Simulation and measurement results demonstrate that the proposed detector enables accurate CUT-related parameter estimation and target detection while 
    achieving empirically stable false-alarm control under the considered calibration procedure. 
    Computational efficiency comparisons further verify its practical feasibility.
\end{enumerate}

The remainder of this paper is organized as follows. 
Section~II introduces the problem formulation and signal model. 
Section~III presents the proposed SA-GLRT-D, including its fast implementation and parameter initialization strategy. 
Section~IV evaluates the proposed methods through Monte Carlo (MC) simulations, measured-data experiments, and computational-efficiency analysis. 
Section~V concludes the paper and discusses future research directions.
\section{Problem Formulation}
\subsection{Signal Model}
Consider an FMCW radar equipped with a uniform linear receive array of \(L\) antennas, 
where each frame contains \(M\) chirps and each chirp has \(N\) fast-time samples. 
The dechirped intermediate-frequency signal at the \(l\)-th receive antenna, denoted by \(\bm{S}_l \in \mathbb{C}^{N \times M}\), can be expressed as
\begin{equation}\label{eq:equ1_Sl}  
    \begin{aligned}
        S_l[n,m] &= \sum_{k=1}^{K} A_{m,k} \exp\!
        \left[\mathrm{j}2\pi \left( f_{\mathrm{b},k} n T_{\mathrm{s}} \!+\! f_{\mathrm{d},k} m T_{\mathrm{PRI}} \right) 
            + \mathrm{j}\phi_{0,k}
        \right] \\
        &\quad \times a_l\left(\theta_k\right) + w_l\left[n,m\right],
    \end{aligned}
\end{equation}
where \(n = 0,1,\ldots,N\!-\!1\) and \(m = 0,1,\ldots,M\!-\!1\) denote the fast-time and slow-time samples, respectively. 
\(A_{m,k}\), \(f_{\mathrm{b},k}\), and \(f_{\mathrm{d},k}\) denote the complex amplitude, beat frequency, and Doppler frequency of the \(k\)-th target in the \(m\)-th chirp, respectively. 
\(T_{\mathrm{s}}\) and \(T_{\mathrm{PRI}}\) denote the sampling period and pulse repetition interval, respectively. 
\(\phi_{0,k}\) denotes the initial phase of the \(k\)-th target, 
and \(w_l\left[n,m\right] \sim \mathcal{CN}\left(0,\sigma^2\right)\) denotes complex Gaussian white noise with zero mean and variance \(\sigma^2\) at the \(l\)-th receive antenna. 
Moreover, \(a_l\left(\theta_k\right)\) denotes the \(l\)-th element of the steering vector corresponding to the \(k\)-th target, which is given by
\begin{equation}\label{eq:equ2}  
    a_l\left(\theta_k\right) = \exp\!\left[-\mathrm{j}2\pi \frac{(l-1)d\sin\theta_k}{\lambda}\right],
\end{equation}
where \(d\) denotes the inter-element spacing, \(\lambda\) denotes the wavelength, and \(\theta_k\) is the angle of arrival of the \(k\)-th target.

To extract target features from radar echoes, the fast Fourier transform (FFT) is typically applied sequentially along the fast-time and slow-time dimensions, 
yielding the range--Doppler information (RDI) for subsequent target detection and feature extraction. The RDI corresponding to \(\bm{S}_l\) after the 2D-FFT, denoted by \(\bm{P}_l\), can be expressed as
\begin{equation}\label{eq:equ3_RDI}  
    \begin{aligned}
        P_l\left[p,q\right] &= 
        \sum_{n=0}^{N-1} \sum_{m=0}^{M-1} S_l\left[n,m\right] \, \omega_{\mathrm{R}}\left[n\right] \exp\!\left(-\mathrm{j}2\pi \frac{pn}{N}\right) \\
        &\quad \times \omega_{\mathrm{V}}\left[m\right] \, \exp\!\left(-\mathrm{j}2\pi \frac{qm}{M}\right),
    \end{aligned}
\end{equation}
where \(p = 0,1,\ldots,N\!-\!1\) and \(q = 0,1,\ldots,M\!-\!1\) denote the range-bin and Doppler-bin indices, respectively.
\(\bm{\omega}_{\mathrm{R}} \in \mathbb{C}^{N \times 1}\) and \(\bm{\omega}_{\mathrm{V}} \in \mathbb{C}^{M \times 1}\) denote the window functions applied along the fast-time and slow-time dimensions, respectively.

\subsection{Hypothesis Testing Framework for CFAR Detection}
FMCW radar target detection is typically performed using CFAR detectors, 
which estimate the local background from sliding-window reference cells under an independent and identically distributed assumption. 
For a prescribed false-alarm probability, the detection threshold is determined according to the Neyman--Pearson (NP) criterion 
\cite{37.Sahed2023ExactClosedFormPfaExpressionsForCAAndGOCFARDetectorsInGammaDistributedRadarClutter}, 
and each CUT is detected via cell-wise hypothesis testing. 
The corresponding decision rule can be expressed as

\begin{equation}\label{eq:equ4}  
    z_{\mathrm{CUT}}^{(r_{\mathrm{D}},\,v_{\mathrm{D}})} \mathop{\gtrless}_{\mathcal{H}_0}^{\mathcal{H}_1} T_{\mathrm{CUT}},
\end{equation}
where \(z_{\mathrm{CUT}}^{(r_{\mathrm{D}},\,v_{\mathrm{D}})} \in \mathbb{R}^{+}\) denotes the observed value of the CUT located at the range-bin index \(r_{\mathrm{D}}\) and Doppler-bin index \(v_{\mathrm{D}}\) in the CFAR detection matrix, 
and \(\mathcal{H}_0\) and \(\mathcal{H}_1\) denote the null hypothesis (target absent) and alternative hypothesis (target present), respectively. 
\(T_{\mathrm{CUT}} \in \mathbb{R}^{+}\) denotes the detection threshold corresponding to the preassigned false alarm probability \(P_{\mathrm{FA}}\), which is defined as
\begin{equation}\label{eq:equ5}  
    T_{\mathrm{CUT}} = \tau \, \xi\left( \left\{z_i\right\}_{i=1}^{N'} \right),
\end{equation}
where \(\tau\) denotes a scaling factor determined by \(P_{\mathrm{FA}}\). 
It should be noted that \(P_{\mathrm{FA}}\) denotes the preassigned false alarm probability, whereas \(P_{\mathrm{fa}}\) denotes the empirical false alarm probability. 
Moreover, \(\xi(\cdot)\) denotes a nonlinear function that maps the reference sample set \(\left\{z_i\right\}_{i=1}^{N'}\) to a nonnegative real value.

\section{Detector Designs}
This section details the proposed SA-GLRT-D, including its three-stage GLRT-based design, 
fast approximation scheme, and initialization and stability enhancement strategy.
\subsection{SA-GLRT-D: A Three-Stage Sidelobe-Aware GLRT-Based Detection Framework}
\subsubsection{Hypothesis Testing Model Formulation}
To formulate local hypothesis testing for the CUT, we define a symmetric observation window centered at 
\((r_{\mathrm{D}}, v_{\mathrm{D}})\) along the fast- and slow-time dimensions as
\begin{equation}\label{eq:equ6_and_7}  
    \begin{aligned}
        \mathscr{P} &= \left\{ \tilde p \in \mathbb{N}\,\middle|\, r_{\mathrm{D}} - \frac{N_{\mathrm{r}} - 1}{2} \le \tilde p \le r_{\mathrm{D}} + \frac{N_{\mathrm{r}} - 1}{2} \right\}, \\
        \mathscr{Q} &= \left\{ \tilde q \in \mathbb{N}\,\middle|\, v_{\mathrm{D}} - \frac{N_{\mathrm{v}} - 1}{2} \le \tilde q \le v_{\mathrm{D}} + \frac{N_{\mathrm{v}} - 1}{2} \right\}.
    \end{aligned}
\end{equation}
where \(\mathscr{P}\) and \(\mathscr{Q}\) denote the fast- and slow-time index sets of the CUT-centered local template, respectively. 
The window sizes \(N_{\mathrm r},N_{\mathrm v}\in\mathbb{N}\) are odd to preserve symmetry, 
and are empirically set to \(N_{\mathrm r}=N_{\mathrm v}=5\) to cover the dominant CUT energy with a compact local region
\cite{13.Zhu2024RobustCFARDetectorBasedOnCLEANForSidelobeSuppression,
30.Cao2021CompressedSensingBasedMultitargetCFARDetectionAlgorithmForFMCWRadar,
35.Zhu2025CTCFARRobustCFARDetector}.

Based on \eqref{eq:equ3_RDI} and \eqref{eq:equ6_and_7}, when only the target located in the CUT is considered, the complex observation at
\(\left(\tilde p,\tilde q\right)\) for the \(l\)-th channel can be simplified as
\begin{equation}\label{eq:equ11}  
    x_{\tilde p,\tilde q}^{(l)} = \alpha \,\mathscr{Y}_{\Delta r,\Delta v}^{(r_{\mathrm{D}},\,v_{\mathrm{D}})}\left[\tilde p,\tilde q\right] 
    + w_l\left[\tilde p,\tilde q\right], \quad l = 1,\ldots,L,
\end{equation}
where \(\alpha \in \mathbb{C}\) denotes the complex amplitude of the target echo, 
and \(\bm{\mathscr{Y}}_{\Delta r,\Delta v}^{(r_{\mathrm{D}},\,v_{\mathrm{D}})}\) denotes the local normalized RDI template of the CUT, which is defined as
\begin{equation}\label{eq:equ8_templateOfCUT}  
    \begin{aligned}
        \bm{\mathscr{Y}}_{\Delta r,\Delta v}^{(r_{\mathrm{D}},\,v_{\mathrm{D}})}\left[\tilde p,\tilde q\right] =&
        \left(\sum_{n=0}^{N-1}\omega_{\mathrm{R}}[n]\,
        \mathrm{e}^{\left[\mathrm{j}2\pi n\left(\hat f_{\mathrm{b}}T_{\mathrm{s}} \!-\! \frac{\tilde p}{N}\right)\right]}\right)\\
        &\times \left(\sum_{m=0}^{M-1}\omega_{\mathrm{V}}[m]\,
        \mathrm{e}^{\left[\mathrm{j}2\pi m\left(\hat f_{\mathrm{d}}T_{\mathrm{PRI}} \!-\! \frac{\tilde q}{M}\right)\right]}\right),
    \end{aligned}
\end{equation}
where \(\hat f_{\mathrm{b}}\) and \(\hat f_{\mathrm{d}}\) denote the beat-frequency and Doppler-frequency estimates of the CUT, respectively, and are defined as
\begin{equation}\label{eq:equ9}  
    \hat f_{\mathrm{b}} = \frac{r_{\mathrm{D}} + \Delta r}{N T_{\mathrm{s}}}, \quad
    \hat f_{\mathrm{d}} = \frac{v_{\mathrm{D}} + \Delta v}{M T_{\mathrm{PRI}}},
\end{equation}
where \(\Delta r,\Delta v\in\mathbb{R}\) denote the range- and Doppler-bin offsets between the CUT indices and the true target frequencies, respectively, with \(\Delta r,\Delta v\in[-0.5,0.5]\).

The CUT-centered local region may contain non-CUT components, 
including interfering targets and sidelobe leakage, which cannot be explained by the CUT-related component alone. 
Rather than explicitly parameterizing these components, we absorb them and the background noise into an effective residual term 
\(\varepsilon_{\tilde p,\tilde q}^{(l)}\). 
Thus, \eqref{eq:equ11} is treated as a single-dominant-target surrogate model, where only the CUT-related component is explicitly modeled. 
For tractability, the residual is approximated as
\(\varepsilon_{\tilde p,\tilde q}^{(l)} \sim \mathcal{CN}(0,c_{\tilde p,\tilde q}\theta)\),
where \(c_{\tilde p,\tilde q}\in\mathbb{R}^{+}\) accounts for spatially varying non-CUT residual power, and \(\theta\) 
denotes the background noise variance estimated by the method in \cite{35.Zhu2025CTCFARRobustCFARDetector}. 
Accordingly, \eqref{eq:equ11} can be reformulated as
\begin{equation}\label{eq:equ10}  
    x_{\tilde p,\tilde q}^{(l)} = \alpha \,\mathscr{Y}_{\Delta r,\Delta v}^{(r_{\mathrm{D}},\,v_{\mathrm{D}})}\left[\tilde p,\tilde q\right] 
    +
    \varepsilon_{\tilde p,\tilde q}^{(l)}, \quad l = 1,\ldots,L,
\end{equation}

Based on the above model, the non-coherent accumulation (NCA) observation corresponding to the CUT, denoted by \(z_{\tilde p,\tilde q}\), is defined as
\begin{equation}\label{eq:equ13}  
    z_{\tilde p,\tilde q} = \sum_{l=1}^{L} \left|x_{\tilde p,\tilde q}^{(l)}\right|^2 .
\end{equation}
The corresponding local window observation vector \(\bm{Z} \in \mathbb{R}^{N_{\mathrm{r}}N_{\mathrm{v}}}\) is then formed by stacking all 
\(z_{i,j}\) for \((i,j) \in \mathscr{P} \times \mathscr{Q}\). 
Accordingly, the hypothesis-testing model within the observation window can be expressed as
\begin{equation}\label{eq:equ14_hypothesisModel}  
    \left\{
        \begin{aligned}
        \mathcal{H}_0(\bm{c}_0): \quad
        & z_{i,j} = \sum_{l=1}^{L} \left| \varepsilon_{i,j}^{(l)} \right|^2, \\
        \mathcal{H}_1(\alpha,\Delta r,\Delta v,\bm{c}_1): \quad
        & z_{i,j} = \sum_{l=1}^{L} \left| x_{i,j}^{(l)} \right|^2,
        \end{aligned}
    \right. 
    \forall\,(i,j)\in\mathscr{P}\times\mathscr{Q},
\end{equation}
where \(\bm{c}_0=\{c_{i,j}^{(0)}\}\) and \(\bm{c}_1=\{c_{i,j}^{(1)}\}\) denote the hypothesis-dependent local variance correction factors for all 
\((i,j)\in\mathscr{P}\times\mathscr{Q}\). 
Rather than representing two physically distinct background-noise variances, 
they provide equivalent statistical descriptions of the residual term under \(\mathcal{H}_0\) and \(\mathcal{H}_1\). 
Because the dominant CUT-related component is explicitly modeled only under \(\mathcal{H}_1\), 
the residual statistics can differ between the two hypotheses, which is captured by the separate correction-factor sets.

Under \(\mathcal{H}_0\), we assume that \(z_{\tilde p,\tilde q}\) are mutually independent and follow Gamma distributions with shape 
\(L\) and scale \(c_{\tilde p,\tilde q}^{(0)}\theta\). This gives the joint probability density function (PDF) of \(\bm{Z}\) as
\begin{equation}\label{eq:equ_15}  
    \begin{aligned}
        f_0(\bm Z;\bm{c}_0)
        =\!\prod_{(i,j)\in\mathscr{P}\times\mathscr{Q}}\! \frac{z_{i,j}^{L-1}}{\Gamma(L)\left(c_{i,j}^{(0)}\theta\right)^L}
        \exp\!\left(-\frac{z_{i,j}}{c_{i,j}^{(0)}\theta}\right).
    \end{aligned}
\end{equation}

Under \(\mathcal{H}_1\), \(x_{\tilde p,\tilde q}^{(l)}\) can be decomposed as
\begin{equation}\label{eq:equ_17}  
    \begin{aligned}
        x_{\tilde p,\tilde q}^{(l)} 
        &= x_{\mathrm{R}\,\tilde p,\tilde q}^{(l)} + \mathrm{j}x_{\mathrm{I}\,\tilde p,\tilde q}^{(l)} \\
        &= \left(m_{\mathrm{R}\,\tilde p,\tilde q} + \varepsilon_{\mathrm{R}\,\tilde p,\tilde q}^{(l)}\right)
        + \mathrm{j}\!\left(m_{\mathrm{I}\,\tilde p,\tilde q} + \varepsilon_{\mathrm{I}\,\tilde p,\tilde q}^{(l)}\right),
    \end{aligned}
\end{equation}
where the subscripts \(\mathrm{R}\) and \(\mathrm{I}\) denote the real and imaginary components, respectively, of 
\(x_{\tilde p,\tilde q}^{(l)}\), 
\(\alpha \mathscr{Y}_{\Delta r,\Delta v}^{(r_{\mathrm D},v_{\mathrm D})}[\tilde p,\tilde q]\), 
and \(\varepsilon_{\tilde p,\tilde q}^{(l)}\).

Hence, \(x_{\mathrm{R},\tilde p,\tilde q}^{(l)}\) and \(x_{\mathrm{I},\tilde p,\tilde q}^{(l)}\) follow Gaussian distributions with means 
\(m_{\mathrm{R},\tilde p,\tilde q}\) and \(m_{\mathrm{I},\tilde p,\tilde q}\), respectively, and variance \(c_{\tilde p,\tilde q}^{(1)}\theta/2\). 
The corresponding NCA observation \(z_{\tilde p,\tilde q}\) is then given by
\begin{equation}\label{eq:equ18}  
    \begin{aligned}
        z_{\tilde p,\tilde q}
        \!=\! \sum_{l=1}^{L}\left|x_{\tilde p,\tilde q}^{(l)}\right|^2
        \!=\! \sum_{l=1}^{L}\left[ \left(x_{\mathrm{R}\,\tilde p,\tilde q}^{(l)}\right)^2 \!+\! \left(x_{\mathrm{I}\,\tilde p,\tilde q}^{(l)}\right)^2 \right],
        \quad z_{\tilde p,\tilde q}\ge 0.
    \end{aligned}
\end{equation}

To characterize the distribution of \(z_{\tilde p,\tilde q}\), we introduce the normalized real and imaginary components of \(x_{\tilde p,\tilde q}^{(l)}\). 
For \(a\in\{\mathrm{R},\mathrm{I}\}\), define

\begin{equation}\label{eq:normalized_component}
    u_{a\,\tilde p,\tilde q}^{(l)}
    =
    \sqrt{\frac{2}{c_{\tilde p,\tilde q}^{(1)}\theta}}\,
    x_{a\,\tilde p,\tilde q}^{(l)} .
\end{equation}
Accordingly, \(u_{a\,\tilde p,\tilde q}^{(l)}\) follows a Gaussian distribution with mean 
\(\sqrt{2/(c_{\tilde p,\tilde q}^{(1)}\theta)}\,m_{a\,\tilde p,\tilde q}\) and unit variance, for \(a\in\{\mathrm{R},\mathrm{I}\}\).

Based on the normalized components, the corresponding normalized energy \(y_{\tilde p,\tilde q}^{(l)}\) is defined as
\begin{equation}\label{eq:equ23}
    y_{\tilde p,\tilde q}^{(l)}
    =
    \left(u_{\mathrm{R}\,\tilde p,\tilde q}^{(l)}\right)^2
    +
    \left(u_{\mathrm{I}\,\tilde p,\tilde q}^{(l)}\right)^2
    =
    \frac{2}{c_{\tilde p,\tilde q}^{(1)}\theta}
    \left|x_{\tilde p,\tilde q}^{(l)}\right|^2,
    \quad
    y_{\tilde p,\tilde q}^{(l)} \ge 0 .
\end{equation}

Since \(u_{\mathrm{R}\,\tilde p,\tilde q}^{(l)}\) and 
\(u_{\mathrm{I}\,\tilde p,\tilde q}^{(l)}\) are independent unit-variance Gaussian variables, 
\(y_{\tilde p,\tilde q}^{(l)}\) follows a noncentral chi-square distribution with two degrees of freedom, i.e.,
\(y_{\tilde p,\tilde q}^{(l)} \sim \chi_2^2(\lambda_{\tilde p,\tilde q}^{(l)})\). 
The noncentrality parameter is given by
\begin{equation}\label{eq:equ25}
    \lambda_{\tilde p,\tilde q}^{(l)}
    =
    \frac{2}{c_{\tilde p,\tilde q}^{(1)}\theta}
    \left|
    \alpha \mathscr{Y}_{\Delta r,\Delta v}^{(r_{\mathrm{D}},v_{\mathrm{D}})}
    [\tilde p,\tilde q]
    \right|^2 .
\end{equation}
Combining \eqref{eq:equ18} and \eqref{eq:equ23}, the normalized accumulated energy over \(L\) receive antennas is given by
\begin{equation}\label{eq:equ26}
    y_{\tilde p,\tilde q} 
    =
    \sum_{l=1}^{L} y_{\tilde p,\tilde q}^{(l)}
    =
    \frac{2}{c_{\tilde p,\tilde q}^{(1)}\theta}\,
    z_{\tilde p,\tilde q},
    \quad
    y_{\tilde p,\tilde q} \ge 0 .
\end{equation}
Then, \(y_{\tilde p,\tilde q}\) follows a noncentral chi-square distribution with \(2L\) degrees of freedom,
i.e., \(y_{\tilde p,\tilde q}\sim\chi_{2L}^{2}(\lambda_{\tilde p,\tilde q})\), where
\begin{equation}\label{eq:equ27}
    \lambda_{\tilde p,\tilde q}
    =
    \frac{2L}{c_{\tilde p,\tilde q}^{(1)}\theta}
    \left|
    \alpha \mathscr{Y}_{\Delta r,\Delta v}^{(r_{\mathrm D},v_{\mathrm D})}
    [\tilde p,\tilde q]
    \right|^2 .
\end{equation}


To characterize the PDF of \(z_{\tilde p,\tilde q}\) under \(\mathcal{H}_1\), 
we first introduce an auxiliary random variable
\(y=\sum_{l=1}^{2L}x_l^2\), where \(\{x_l\}_{l=1}^{2L}\) are independent real Gaussian random variables with means \(\mu_l\) and unit variance. 
Then, \(y\) follows a noncentral chi-square distribution with \(2L\) degrees of freedom and noncentrality parameter 
\(\lambda=\sum_{l=1}^{2L}\mu_l^2\), i.e., \(y\sim\chi_{2L}^2(\lambda)\). 
Accordingly, the PDF of \(y\) is given by
\begin{equation}\label{eq:equ40}
    f_y\left( y \right)
    =
    \frac{1}{2}
    \exp\!\left(-\frac{y+\lambda}{2}\right)
    \left(\frac{y}{\lambda}\right)^{\frac{L-1}{2}}
    I_{L-1}\!\left(\sqrt{\lambda y}\right),
    \quad y\ge 0,
\end{equation}
where \(I_{L-1}(\cdot)\) denotes the modified Bessel function of the first kind.

Using the scaling relation in \eqref{eq:equ26}, 
the joint PDF of \(\bm{Z}\) under \(\mathcal{H}_1\) is obtained by change of variables as given in \eqref{eq:equ45_Z_pdfH1}.
\begin{figure*}[!b]  
\centering \hrulefill
    \begin{align}
        \label{eq:equ45_Z_pdfH1}  
        f_1\!\left( \bm{Z};
        \bm{\lambda}, \bm{c}_1 \right)
        &=\!
        \prod_{(i,j)\in\mathscr{P}\times\mathscr{Q}}\!
        \frac{1}{c_{i,j}^{(1)}\theta}\exp\!\left(-\frac{z_{i,j}}{c_{i,j}^{(1)}\theta} - \frac{\lambda_{i,j}}{2}\right)
        \left(\frac{2z_{i,j}}{c_{i,j}^{(1)}\theta \lambda_{i,j}}\right)^{\frac{L-1}{2}}
        I_{L-1}\!\left(\sqrt{\frac{2\lambda_{i,j}z_{i,j}}{c_{i,j}^{(1)}\theta}}\right).
    \end{align}
\vspace*{-0.9cm}
\end{figure*}
For notational simplicity, \(\lambda_{\tilde p,\tilde q}\) and \(\bm{\lambda}\) hereafter denote
\(\lambda_{\tilde p,\tilde q}(\alpha,\Delta r,\Delta v,c_{\tilde p,\tilde q}^{(1)})\)
and
\(\bm{\lambda}(\alpha,\Delta r,\Delta v,\bm{c}_1)\), respectively, where
\(\bm{\lambda}=\{\lambda_{\tilde p,\tilde q}\}_{(\tilde p,\tilde q)\in\mathscr{P}\times\mathscr{Q}}\in\mathbb{R}^{N_{\mathrm r}N_{\mathrm v}}\).
The joint PDFs of \(\bm{Z}\) under \(\mathcal{H}_0\) and \(\mathcal{H}_1\) are summarized in \eqref{eq:equ47_joint_pdf}.
\begin{figure*}[!b]  
\centering \hrulefill
    \begin{align}
        \label{eq:equ47_joint_pdf}  
        \left\{
            \begin{aligned}
            \mathcal{H}_0\!\left( \bm{c}_0 \right):\quad&
            f_0\!\left( \bm{Z};\bm{c}_0 \right)
            =\!
            \prod_{(i,j)\in\mathscr{P}\times\mathscr{Q}}\!
            \frac{1}{\Gamma(L)\left(c_{i,j}^{(0)}\theta\right)^L}
            z_{i,j}^{L-1}
            \exp\!\left(
                -\frac{z_{i,j}}{c_{i,j}^{(0)}\theta}
            \right), \\
            \mathcal{H}_1\!\left( \alpha,\Delta r,\Delta v,\bm{c}_1 \right):\quad&
            f_1\!\left( \bm{Z};\bm{\lambda},\bm{c}_1 \right)
            =\!
            \prod_{(i,j)\in\mathscr{P}\times\mathscr{Q}}\!
            \frac{1}{c_{i,j}^{(1)}\theta}
            \exp\!\left(-\frac{z_{i,j}}{c_{i,j}^{(1)}\theta} -\frac{\lambda_{i,j}}{2}\right)
            \left(\frac{2z_{i,j}}{c_{i,j}^{(1)}\theta \lambda_{i,j}}\right)^{\frac{L-1}{2}}
            I_{L-1}\!\left(\sqrt{\frac{2\lambda_{i,j}z_{i,j}}{c_{i,j}^{(1)}\theta}}\right).
            \end{aligned}
        \right.
    \end{align}
\vspace*{-0.9cm}
\end{figure*}
\subsubsection{Construction of the GLRT-Based Decision Statistic}
%
%
%
In this binary hypothesis-testing problem, both hypotheses involve multiple unknown parameters, 
which makes the optimal NP likelihood-ratio test analytically intractable without prior parameter knowledge. 
As a practical alternative, we consider the GLRT, in which the unknown parameters under 
\(\mathcal{H}_0\) and \(\mathcal{H}_1\) are estimated from the observed data and substituted into the likelihood ratio. 
The formal GLRT statistic is given by
%
%
\begin{equation}\label{eq:equ48}  
    \Lambda(\bm{Z})=
    \frac{\displaystyle\max_{\Delta r,\Delta v}\max_{\alpha}\max_{\bm{c}_1}f_1\!\left( \bm{Z}; \bm{\lambda}, \bm{c}_1 \right)}
    {\displaystyle\max_{\bm{c}_0}f_0\!\left( \bm{Z}; \bm{c}_0 \right)}.
\end{equation}
Since the joint maximization over \((\rho,\Delta r,\Delta v,\bm{c}_1)\) is analytically intractable, 
we implement \eqref{eq:equ48} using a plug-in GLRT strategy, 
where the unknown parameters are estimated through block-wise maximization and then substituted into the likelihood ratio. 
The resulting statistic can therefore be regarded as an approximate GLRT statistic, 
while preserving the hypothesis-testing structure of the original formulation.

\paragraph{Estimation of \(\hat{\bm{c}}_{0}\)}  
Under \(\mathcal{H}_0\), the joint log-likelihood function of \(f_0\) can be expressed as
\begin{equation}\label{eq:equ49}  
    \ell_0\!\left(\bm{c}_0 \right) =
    \!\sum_{(i,j)\in\mathscr{P}\times\mathscr{Q}}\!
    \ln \!\left[\frac{1}{\Gamma(L)\left(c_{i,j}^{(0)}\theta\right)^L}z_{i,j}^{L-1}
    \mathrm{e}^{-\frac{z_{i,j}}{c_{i,j}^{(0)}\theta}}\right].
\end{equation}
Therefore, the pointwise closed-form estimate of \(\hat{\bm{c}}_0\) can be obtained as
\begin{equation}\label{eq:equ50}  
    \hat{\bm{c}}_0
    =\left\{\frac{z_{i,j}}{\mu_z}\right\}_{(i,j)\in\mathscr{P}\times\mathscr{Q}},
\end{equation}
where \(\mu_z\) denotes the mean of the background noise and is given by \(\mu_z=L\theta\) \cite{35.Zhu2025CTCFARRobustCFARDetector}.

\paragraph{Estimation of \(\hat{\bm{c}}_{1}\)}  
Under \(\mathcal{H}_1\), the joint log-likelihood function of \(f_1\) can be expressed as
\begin{equation}\label{eq:equ51}  
    \ell_1\!\left( \alpha,\Delta r,\Delta v,\bm{c}_1 \right)=
    \!\sum_{(i,j)\in\mathscr{P}\times\mathscr{Q}}\!
    \ln \!\left[f_1\!\left( z_{i,j}; \lambda_{i,j}, c_{i,j}^{(1)} \right)\right].
\end{equation}

To facilitate the inner-layer optimization with respect to \(\bm{c}_1\), 
we fix \(\alpha\), \(\Delta r\), and \(\Delta v\). The first-order derivative of \(\ell_1\) with respect to \(c_{i,j}^{(1)}\) is then given by
\begin{equation}\label{eq:equ52}  
    \frac{{\mathrm d}\ell_1}{{\mathrm d}c_{i,j}^{(1)}}
    =
    - \frac{1}{c_{i,j}^{(1)}}
    + \frac{1}{\left(c_{i,j}^{(1)}\right)^2}
    \left(
        \frac{z_{i,j}}{\theta}
        + \frac{\beta_{i,j}}{2}
    \right)
    + \frac{{\mathrm d}\ln I_{L-1}\!\left( t_{i,j} \right)}{{\mathrm d}c_{i,j}^{(1)}},
\end{equation}
where \(\beta_{i,j}\) and \(t_{i,j}\) are defined as
\begin{equation}\label{eq:equ53}  
    \beta_{i,j} =
    \frac{2L}{\theta}
    \left| \alpha \mathscr{Y}_{\Delta r,\Delta v}^{\left( r_{\mathrm D}, v_{\mathrm D} \right)}[i,j] \right|^2,
    \quad
    t_{i,j} =
    \frac{1}{c_{i,j}^{(1)}}
    \sqrt{\frac{2\beta_{i,j}z_{i,j}}{\theta}}.
\end{equation}

The first two terms of \({\mathrm d}\ell_1/{\mathrm d}c_{i,j}^{(1)}\) are directly computable, 
while the last term depends on the logarithmic derivative of \(I_v(x)\). 
By combining the derivative expression of \(I_v(x)\) with the recurrence relation between \(I_{v-1}(x)\) and \(I_{v+1}(x)\), 
as given in \eqref{eq:equ55_IvDerivative} and \eqref{eq:equ56_IvDiff}, 
the first-order derivative of \(\ln I_v(x)\) is obtained as
\begin{figure*}[!b]  
\centering 
    \begin{align}
        \label{eq:equ55_IvDerivative}  
        I_v'(x)
        =\frac{1}{2}\sum_{i=0}^{\infty}\frac{2i+v}{i!\,\Gamma(i+v+1)}\left(\frac{x}{2}\right)^{2i+v-1}
        =\frac{1}{2}\sum_{i=0}^{\infty}\left[\frac{i}{i!\,\Gamma(i+v+1)}+\frac{i+v}{i!\,\Gamma(i+v+1)}\right]\left(\frac{x}{2}\right)^{2i+v-1}
        =\frac{1}{2}\left[I_{v+1}(x) + I_{v-1}(x)\right].
    \end{align}
\vspace*{-0.9cm}
\end{figure*}
\begin{figure*}[!b]  
\centering 
    \begin{align}
        \label{eq:equ56_IvDiff}  
        I_{v-1}(x)-I_{v+1}(x)
        =\sum_{i=0}^{\infty}\frac{1}{i!\Gamma(i+v)}\left(\frac{x}{2}\right)^{2i+v-1} -\sum_{i=0}^{\infty}\frac{1}{i!\Gamma(i+v+2)}\left(\frac{x}{2}\right)^{2i+v+1}
        =\frac{2v}{x} I_v(x).
    \end{align}
\vspace*{-0.9cm}
\end{figure*}
\begin{equation}\label{eq:equ57}  
    \frac{d}{dx}\ln I_v(x)
    = \frac{{I_{v + 1}}\left( x \right) + {I_{v - 1}}\left( x \right)}{{2{I_v}\left( x \right)}} 
    = \frac{{{I_{v + 1}}\left( x \right)}}{{{I_v}\left( x \right)}} + \frac{v}{x}.
\end{equation}

Substituting \(v=L-1\) and \(x=t_{i,j}\) into \eqref{eq:equ57} and applying the chain rule to \eqref{eq:equ52}, 
the stationary-point condition with respect to \(c_{i,j}^{(1)}\) is given by
\begin{equation}\label{eq:equ60}  
    Lc _{i,j}^{\left( 1 \right)} = 
    \frac{{{z_{i,j}}}}{\theta } 
    +\frac{{{\beta _{i,j}}}}{2} 
    -{t_{i,j}}c _{i,j}^{\left( 1 \right)}\frac{{{I_L}\left( {{t_{i,j}}} \right)}}{{{I_{L - 1}}\left( {{t_{i,j}}} \right)}}.
\end{equation}

Since \eqref{eq:equ60} is an implicit nonlinear equation in \(c_{i,j}^{(1)}\), a closed-form solution is difficult to obtain. 
To solve it numerically, \eqref{eq:equ60} is reformulated as a fixed-point equation with respect to \(t_{i,j}\), yielding
\begin{equation}\label{eq:equ64}
    t_{i,j}^{(n+1)} = \frac{L}{\mathcal{A}_{i,j} - r\!\left( t_{i,j}^{(n)} \right)} .
\end{equation}
where \(t_{i,j}^{(n)}\) denotes the estimate at the \(n\)-th iteration, 
\(\mathcal{A}_{i,j}\) and \(r\!\left( t_{i,j}^{(n)} \right)\) are respectively defined as
\begin{equation}\label{eq:equ63}
    \mathcal{A}_{i,j} = \frac{z_{i,j}/\theta + \beta_{i,j}/2}{k_{i,j}},
    \quad
    r\!\left( t_{i,j}^{(n)} \right) = 
    \frac{I_L\!\left( t_{i,j}^{(n)} \right)}
    {I_{L-1}\!\left( t_{i,j}^{(n)} \right)} .
\end{equation}
According to \eqref{eq:equ53}, \(k_{i,j}\) is given by
\begin{equation}\label{eq:equ61}  
    k_{i,j} 
    = \sqrt{\frac{2\beta_{i,j}z_{i,j}}{\theta}}
    = c_{i,j}^{(1)}t_{i,j}.
\end{equation}
Once the iteration converges to \(\hat t_{i,j}\), the pointwise estimate \(\hat c_{i,j}^{(1)}\) can be obtained from \eqref{eq:equ61}, 
and the vector estimate \(\hat{\bm{c}}_1\) can be obtained.

\paragraph{Estimation of \(\hat{\alpha}\)}  
Under the NCA structure, \(\alpha\) enters \(\lambda_{i,j}(\cdot)\) only through \(|\alpha|^2\). 
Therefore, the maximization over \(\alpha\) can be equivalently reformulated as an optimization over the nonnegative scalar \(\rho=|\alpha|^2\), i.e.,
\begin{equation}\label{eq:equ65}  
    \max_{\alpha \in \mathbb{C}}\ell_1\!\left( \alpha,\Delta r,\Delta v,\hat{\bm{c}}_1 \right)
    \equiv
    \max_{\rho \ge 0}\ell_1\!\left( \rho,\Delta r,\Delta v,\hat{\bm{c}}_1 \right).
\end{equation}

To facilitate numerical optimization over \(\rho\), 
\(\lambda_{i,j}\) can be rewritten according to \eqref{eq:equ27} as a linear function of \(\rho\), 
i.e., \(\lambda_{i,j}=\beta_{i,j}'\rho\), 
where \(\beta_{i,j}'\) denotes the corresponding coefficient defined as
\begin{equation}\label{eq:equ67}  
    \beta_{i,j}'
    =
    \frac{2L}{\theta \hat{c}_{i,j}^{(1)}}
    \left|
        \mathscr{Y}_{\Delta r,\Delta v}^{\left( r_{\mathrm D}, v_{\mathrm D} \right)}\left[i,j\right]
    \right|^2.
\end{equation}
Meanwhile, the argument of the logarithmic Bessel term in \(\ell_1\) can also be expressed as a function of \(\rho\), denoted by \(t_{i,j}'(\rho)\), i.e.,
\begin{equation}\label{eq:equ68}  
    t_{i,j}'(\rho)
    =\sqrt{\frac{2z_{i,j}}{\theta \hat{c}_{i,j}^{(1)}}\lambda_{i,j}}
    =\sqrt{\frac{2\beta_{i,j}'z_{i,j}}{\theta \hat{c}_{i,j}^{(1)}}\rho}.
\end{equation}
With \((\Delta r,\Delta v)\) fixed and \(\hat{\bm{c}}_1\) substituted, the inner-layer estimation of \(\rho\) is performed by maximizing \(\ell_1\). 
The corresponding objective function, denoted by \(\mathcal{J}_\rho\), is defined in \eqref{eq:equ69_rhoObjFun}.
\begin{figure*}[!b]  
\centering \hrulefill
\begin{equation}
    \begin{aligned}
        \label{eq:equ69_rhoObjFun}  
        \mathcal{J}_\rho\! 
        = \!\sum_{(i,j)\in\mathscr{P}\times\mathscr{Q}}\!
        \left[
            \ln \frac{1}{\theta \hat{c}_{i,j}^{(1)}} - \frac{z_{i,j}}{\theta \hat{c}_{i,j}^{(1)}} + \frac{L-1}{2}
            \ln \left(\frac{2z_{i,j}}{\theta \hat{c}_{i,j}^{(1)}\beta_{i,j}'}\right)
            - \frac{\beta_{i,j}'}{2}\rho - \frac{L-1}{2}\ln \rho + \ln I_{L-1}\!\left( t_{i,j}'(\rho) \right)
        \right].
    \end{aligned}
\end{equation}
\vspace*{-0.9cm}
\end{figure*}

Accordingly, the optimization problem for estimating \(\rho\) is formulated as
\begin{equation}\label{eq:equ70}  
    \hat{\rho}\!
    = \arg\max_{\rho \ge 0}
    \mathcal{J}_\rho\!\left(\rho;\Delta r,\Delta v,\hat{\bm{c}}_1\right).
\end{equation}

To determine the optimal CUT power parameter \(\hat{\rho}\), 
we differentiate \(\mathcal{J}_\rho\) with respect to \(\rho\) and use the result in \eqref{eq:equ57} to handle the Bessel-function derivative. 
Accordingly, \(\hat{\rho}\) can be determined from the following one-dimensional implicit equation as
\begin{equation}\label{eq:equ74}  
    \rho\!\sum_{(i,j)\in\mathscr{P}\times\mathscr{Q}}\!
    \beta_{i,j}'
    = \!\sum_{(i,j)\in\mathscr{P}\times\mathscr{Q}}\!
    t_{i,j}'(\rho)\frac{I_L\!\left( t_{i,j}'\left(\rho\right) \right)}{I_{L-1}\!\left( t_{i,j}'(\rho) \right)}.
\end{equation}

Since \eqref{eq:equ74} is an implicit nonlinear equation, a closed-form solution is difficult to obtain. 
To solve it numerically, an iterative scheme similar to that for \(\bm{c}_1\) is adopted, yielding
\begin{equation}\label{eq:equ75}  
    \rho^{(n+1)}
    =
    \frac{\displaystyle
        \!\sum_{(i,j)\in\mathscr{P}\times\mathscr{Q}}\!
        t_{i,j}'\!\left( \rho^{(n)} \right)
        \frac{I_L\!\left( t_{i,j}'\!\left( \rho^{(n)} \right) \right)}{I_{L-1}\!\left( t_{i,j}'\!\left( \rho^{(n)} \right) \right)}}{
        \displaystyle
        \!\sum_{(i,j)\in\mathscr{P}\times\mathscr{Q}}\!
        \beta_{i,j}'},
\end{equation}
where \(\rho^{(n)}\) denotes the estimate of the power parameter at the \(n\)-th iteration. Once the iteration converges, the estimate \(\hat{\rho}\) is obtained.

\paragraph{Estimation of \(\left(\Delta \hat r,\Delta \hat v\right)\)}  
The objective function \(\mathcal{J}_\Delta\) with respect to \((\Delta r,\Delta v)\) is defined in \eqref{eq:equ76_deltaObjFun}.
\begin{figure*}[!b]  
\centering 
\begin{equation}
    \begin{aligned}
        \label{eq:equ76_deltaObjFun}  
        \mathcal{J}_\Delta =
        \!\sum_{(i,j)\in\mathscr{P}\times\mathscr{Q}}\!
        \Bigg[
            -\frac{\upsilon \mathcal{Y}_{i,j}(\Delta r,\Delta v)}{\mathcal{B}_{i,j}}
            -\frac{L-1}{2}\ln \mathcal{Y}_{i,j}(\Delta r,\Delta v)
            +\ln I_{L-1}\!\left(\frac{2\sqrt{\upsilon z_{i,j}\mathcal{Y}_{i,j}(\Delta r,\Delta v)}}{\mathcal{B}_{i,j}}\right)
        \Bigg].
    \end{aligned}
\end{equation}
\vspace*{-0.9cm}
\end{figure*}
Here, \(\mathcal{B}_{i,j}\), \(\upsilon\), and \(\mathcal{Y}_{i,j}(\Delta r,\Delta v)\) are defined as
\begin{equation}\label{eq:equ77}  
    \mathcal{B}_{i,j} = \theta \hat{c}_{i,j}^{(1)},
    \quad
    \upsilon = L\hat{\rho},
\end{equation}
and
\begin{equation}\label{eq:equ78}  
    \mathcal{Y}_{i,j}(\Delta r,\Delta v)
    =\left|\mathscr{Y}_{\Delta r,\Delta v}^{\left( r_{\mathrm D}, v_{\mathrm D} \right)}[i,j]\right|^2.
\end{equation}

Accordingly, the optimization problem with respect to \((\Delta r,\Delta v)\) can be formulated as
\begin{equation}\label{eq:equ79}  
    \left(\Delta \hat r,\Delta \hat v\right)
    =\mathop{\arg\max}\limits_{(\Delta r,\Delta v)\in[-0.5,\,0.5]}
    \mathcal{J}_\Delta\!\left(\Delta r,\Delta v;\hat{\rho},\hat{\bm{c}}_1\right).
\end{equation}
To balance complexity and accuracy, \(\mathcal{J}_\Delta\) is maximized via a two-stage grid search. 
A coarse search is first performed over a \(1\times1\) region centered at \((r_{\mathrm D},v_{\mathrm D})\) with step size \(\vartheta_1\). 
The resulting estimate then centers a \(2\vartheta_1\times2\vartheta_1\) fine-search window with step size \(\vartheta_2\), producing \((\Delta\hat r,\Delta\hat v)\).
\subsubsection{GLRT-Based Detection, Threshold Setting, and CLEAN-Inspired Sidelobe Suppression}
In summary, by substituting the estimates \(\hat{\bm{c}}_0\), \(\hat{\bm{c}}_1\), \(\hat{\rho}\), and \((\Delta \hat r,\Delta \hat v)\) 
into the GLRT in \eqref{eq:equ48}, the resulting GLRT statistic for target detection can be defined as
\begin{equation}\label{eq:equ80}  
    T\left(\bm{Z}\right)
    = 2\ln \Lambda(\bm{Z})
    = 2\ln \left[\frac{
        f_1\!\left(\bm{Z};\bm{\lambda}\!\left( \hat{\rho},\Delta \hat r,\Delta \hat v,\hat{\bm{c}}_1 \right),\hat{\bm{c}}_1\right)}
        {f_0\!\left( \bm{Z};\hat{\bm{c}}_0 \right)}\right].
\end{equation}
The corresponding decision rule can be written as
\(T(\bm{Z}) \mathop{\gtrless}_{\mathcal H_0}^{\mathcal H_1} \eta\),
where \(\eta\) denotes the detection threshold. For a prescribed \(P_{\mathrm{FA}}\), 
\(\eta\) is determined by the cumulative distribution function (CDF) of \(T(\bm{Z})\) under \(\mathcal H_0\), i.e.,
\begin{equation}\label{eq:equ82}
    P\!\left( T(\bm{Z}) \le \eta \mid \mathcal H_0 \right)
    = 1 - P_{\mathrm{FA}}
    = F_{T\mid \mathcal H_0}(\eta),
\end{equation}
where \(F_{T\mid \mathcal H_0}(\cdot)\) denotes the CDF of \(T(\bm{Z})\) under \(\mathcal H_0\).

Since \(T(\bm{Z})\) involves nonlinear operations, including parameter maximization and grid search, the CDF 
\(F_{T\mid\mathcal H_0}(\cdot)\) is generally unavailable in closed form. 
Thus, the simulation threshold is calibrated by MC simulation 
\cite{16.Wang2024JointDetectionAndDelayDopplerEstimationAlgorithmsForMIMORadars}. 
Specifically, \(N_{\mathrm{MC}}=100/P_{\mathrm{FA}}\) independent samples are generated under \(\mathcal H_0\), 
yielding an approximate \(10\%\) relative standard deviation for the tail-probability estimate. 
The threshold is estimated as the empirical \((1-P_{\mathrm{FA}})\)-quantile of 
\(\{T(\bm{Z}^{(n)})\}_{n=1}^{N_{\mathrm{MC}}}\), i.e.,
\begin{equation}\label{eq:equ84}
    \eta 
    = F_{T\mid\mathcal H_0}^{-1}(1-P_{\mathrm{FA}})
    \approx T_{(\lceil N_{\mathrm{MC}}(1-P_{\mathrm{FA}})\rceil)},
\end{equation}
where \(T_{(k)}\) denotes the \(k\)-th order statistic, and \(\lceil \cdot \rceil\) denotes the ceiling function.
For measured-data experiments, target-free observations collected by pointing the radar toward the sky are used as \(\mathcal H_0\) samples, 
and the thresholds are obtained using the same empirical-quantile procedure.

In the \(k\)-th target detection iteration, the cell corresponding to the global maximum of the CFAR detection matrix \(\bm{P}_{\mathrm{NCA}}^{(k)}\) is selected as the CUT, denoted by
\begin{equation}\label{eq:equ88}  
    z_{\mathrm{CUT}}^{(r_{\mathrm D},v_{\mathrm D})}
    = \max\!\left( \bm{P}_{\mathrm{NCA}}^{(k)} \right), \quad k = 0,1,2,\ldots .
\end{equation}

Subsequently, the local region centered at the selected CUT \((r_{\mathrm D},v_{\mathrm D})\) is extracted from \(\bm{P}_{\mathrm{NCA}}^{(0)}\) to form \(\bm{Z}\). 
If \(T(\bm{Z})\ge\eta\), the CUT is declared as a target. 
A CLEAN-inspired procedure is then used to suppress window-induced mainlobe broadening and sidelobe-induced false alarms by removing the detected target 
and its associated sidelobes. 
Using \(\hat{\rho}\) and \((\Delta\hat r,\Delta\hat v)\), the corresponding NCA RDI contribution 
\(\bm{P}_{\mathrm{CUT}}\in\mathbb{R}^{N\times M}\) is reconstructed as
\begin{equation}\label{eq:equ89}  
    P_{\mathrm{CUT}}\left[i,j\right]
    =
    \begin{cases}
        \hat{\rho}L
        \left|\mathscr{Y}_{\Delta \hat r,\Delta \hat v}^{\left( r_{\mathrm D},v_{\mathrm D} \right)}[i,j]\right|^2,
        &
        \begin{aligned}[c]
            i \in \mathscr{P}\big|_{N_{\mathrm r}=N_{\mathrm r}'}, \\
            j \in \mathscr{Q}\big|_{N_{\mathrm v}=N_{\mathrm v}'},
        \end{aligned}
        \\[4pt] 
        0, & \text{otherwise},
    \end{cases}
\end{equation}
where \(N_{\mathrm r}'\) and \(N_{\mathrm v}'\) specify the local reconstruction-window size. 
The reconstruction window is chosen slightly larger than the detection reference window to cover the CUT mainlobe and sidelobes. 
The CFAR detection matrix is then updated as
\begin{equation}\label{eq:equ90}  
    \bm{P}_{\mathrm{NCA}}^{(k+1)}
    = \bm{P}_{\mathrm{NCA}}^{(k)} - g_{\mathrm C}\bm{P}_{\mathrm{CUT}},
\end{equation}
where \(g_{\mathrm C}\) denotes the gain factor of the CLEAN-inspired, 
which is used to control the suppression strength of the leakage components in the target spectrum.

By iterating ``parameter estimation--target discrimination--spectrum suppression'', 
SA-GLRT-D sequentially detects targets and suppresses spectral-leakage-induced false alarms. 
The iteration stops when no candidate cell in the updated \(\bm{P}_{\mathrm{NCA}}^{(k+1)}\) satisfies the threshold criterion. 
The overall workflow is summarized in \textbf{Algorithm \ref{alg:algorithm1}}.
\begin{algorithm}[t]
    \caption{SA-GLRT-D.}\label{alg:algorithm1}
    \begin{spacing}{0.97}
    \begin{algorithmic}[1]
        \Statex \textbf{Input\hspace{0.5cm}:} $\left\{\bm{S}_l\right\}_{l=1}^{L},\ P_{\mathrm{FA}}$
        \Statex \textbf{Initialize:} $\bm{N}_\mathrm{G} \gets \operatorname{NoiseEst}(\bm{S}_l)$
        \Statex \hspace{1.45cm} $\bm{P}_{\mathrm{NCA}}^{(0)},\ \bm{Z}_{\mathrm{NCA}} \gets \sum_{l=1}^{L} |\bm{S}_l|^2$
        \Statex \hspace{1.45cm} $\mathcal{T} \gets \emptyset,\ k \gets 0$
        \Statex \textbf{Output\hspace{0.25cm}:} Target list $\mathcal{T}$

        \While{true}
            \State $(r_\mathrm{D}, v_\mathrm{D}) \gets \arg\max\!\bigl(\bm{P}_{\mathrm{NCA}}^{(k)}\bigr)$
            \State $z_{\mathrm{CUT}}^{(r_\mathrm{D},v_\mathrm{D})} \gets \bm{P}_{\mathrm{NCA}}^{(k)}(r_\mathrm{D},v_\mathrm{D})$
            \State $\hat{\bm{c}}_0,\hat{\bm{c}}_1,\hat{\rho},\Delta\hat{r},\Delta\hat{v}\gets \operatorname{ParaEst}\!\left(\bm{Z}_{\mathrm{NCA}}, z_{\mathrm{CUT}}^{(r_\mathrm{D},v_\mathrm{D})}, \bm{N}_\mathrm{G}\right)$
            \State $T \gets \operatorname{GLRT}\!\left(\hat{\bm{c}}_0,\hat{\bm{c}}_1,\hat{\rho},
            \Delta\hat{r},\Delta\hat{v}, \bm{Z}_{\mathrm{NCA}}, z_{\mathrm{CUT}}^{(r_\mathrm{D},v_\mathrm{D})}\right)$
            \State $\eta \gets \operatorname{ThresholdEst}(P_{\mathrm{FA}}, \hat{\bm{c}}_0, \bm{N}_\mathrm{G})$
            \If{$T > \eta$}
                \State $\bm{P}_{\mathrm{CUT}} \gets \operatorname{Reconstruct}(r_\mathrm{D}, v_\mathrm{D}, \hat{\rho}, \Delta\hat{r}, \Delta\hat{v})$
                \State $\bm{P}_{\mathrm{NCA}}^{(k+1)} \gets \operatorname{CLEAN}(\bm{P}_{\mathrm{NCA}}^{(k)}, \bm{P}_{\mathrm{CUT}}),\quad k \gets k+1$
                \State $\mathcal{T} \gets \mathcal{T} \cup \left\{(r_\mathrm{D}, v_\mathrm{D}, \hat{\rho}, \Delta\hat{r}, \Delta\hat{v})\right\}$
            \Else
                \State \textbf{break}
            \EndIf
        \EndWhile

        \State \textbf{Return:} Target list $\mathcal{T}$
    \end{algorithmic}
    \end{spacing}
\end{algorithm}

\subsection{Fast Approximation Scheme: \upshape SA-GLRT-D\textsubscript{F}}
Since SA-GLRT-D requires frequent evaluations of the modified Bessel function ratio 
\(\mathcal{R}_L(x) \triangleq I_L(x)/I_{L-1}(x)\) when constructing the GLRT statistic and further employs a two-stage grid search to optimize 
\((\Delta \hat r,\Delta \hat v)\), its overall computational burden can be relatively high. 
To improve runtime efficiency, we develop a fast approximation scheme, termed SA-GLRT-D\(_{\mathrm F}\). 
The proposed scheme omits the fine grid-search stage and replaces the direct numerical evaluation of \(\mathcal{R}_L(x)\) 
with a low-order analytical approximation, thereby reducing computational cost while incurring only a modest performance loss.

In particular, based on the power-series expansion of the modified Bessel function, 
if \(I_L(x)\) and \(I_{L-1}(x)\) are both expanded up to the \(x^4\) term, 
a low-order analytical approximation of \(\mathcal{R}_L\left(x\right)\) can be obtained as
\begin{equation}\label{eq:equ92}  
    \begin{aligned}
        \mathcal{R}_L\left(x\right) =
        \frac{x}{2L} \!\times\!
        \frac{1 + \frac{1}{L+1}\left(\frac{x}{2}\right)^2 + \frac{1}{2(L+1)(L+2)}\left(\frac{x}{2}\right)^4 + o(x^6)}
        {1 + \frac{1}{L}\left(\frac{x}{2}\right)^2 + \frac{1}{2L(L+1)}\left(\frac{x}{2}\right)^4 + o(x^6)}.
    \end{aligned}
\end{equation}

When \(x \to 0\), by performing a Taylor expansion of the denominator in \eqref{eq:equ92} and retaining terms up to \(x^4\), 
the low-order approximation of \(\mathcal{R}_L\left(x\right)\) can be simplified as shown in \eqref{eq:equ93}.
\begin{figure*}[!b]  
\centering \hrulefill
\begin{equation}\label{eq:equ93}  
    \begin{aligned}
        \mathcal{R}_L\left(x\right)
        &=\frac{x}{2L}
        \left[1 + \frac{1}{L+1}\left(\frac{x}{2}\right)^2 + \frac{1}{2(L+1)(L+2)}\left(\frac{x}{2}\right)^4 + o(x^6)\right]
        \left[1 - \frac{x^2}{4L} + \frac{L+2}{32L^2(L+1)}x^4 + o(x^6)\right] \\
        &\approx
        \frac{x}{2L}
        \left[1 - \frac{x^2}{4L(L+1)} + \frac{x^4}{8L^2(L+1)(L+2)}\right].
    \end{aligned}
\end{equation}
\vspace*{-0.9cm}
\end{figure*}

When \(x \to \infty\), according to the asymptotic expansion of the modified Bessel function given in Chapter 10 of the DLMF 
\cite{38.NISTDigitalLibraryOfMathematicalFunctions}, \(I_v(x)\) can be expressed as
\begin{equation}\label{eq:equ94}  
    \begin{aligned}
        &I_v(x) \approx \\
        &\frac{\exp(x)}{\sqrt{2\pi x}}
        \left[1 \!-\! \frac{4v^2-1}{8x} \!+\! \frac{\left(4v^2-1\right)\left(4v^2-9\right)}{2!\left(8x\right)^2} 
        \!+\! 
        o\!\left(\frac{1}{x^3}\right)\right].
    \end{aligned}
\end{equation}
Accordingly, \(\mathcal{R}_L\left(x\right)\) can be expressed  as
\begin{equation}\label{eq:equ95}  
    \mathcal{R}_L\left(x\right) \approx
    \frac{1-\frac{\mathcal{C}_L}{x}+\frac{\mathcal{D}_L}{x^2}+o\!\left(\frac{1}{x^3}\right)}
    {1-\frac{\mathcal{C}_{L-1}}{x}+\frac{\mathcal{D}_{L-1}}{x^2}+o\!\left(\frac{1}{x^3}\right)},
\end{equation}
where \(\mathcal{C}_v\) and \(\mathcal{D}_v\), \(v \in \left\{ L-1,\,L \right\}\), are defined as
\begin{equation}\label{eq:equ96_and_97}  
    \mathcal{C}_v = \frac{4v^2-1}{8}, \quad
    \mathcal{D}_v = \frac{\left(4v^2-1\right)\left(4v^2-9\right)}{128}.
\end{equation}
Then, by performing a Taylor expansion of the denominator in \eqref{eq:equ95} and retaining the terms required up to \(x^{-4}\), 
\(\mathcal{R}_L\left(x\right)\) can be approximated as shown in \eqref{eq:equ98}.
\begin{figure*}[!b]  
\centering  
\begin{equation}\label{eq:equ98}  
    \begin{aligned}
        \mathcal{R}_L\left(x\right)
        =
        \left(1 - \frac{\mathcal{C}_L}{x}+\frac{\mathcal{D}_L}{x^2}\right)
        \left[1 + \frac{\mathcal{C}_{L-1}}{x}-\frac{\mathcal{D}_{L-1}}{x^2} + \left(-\frac{\mathcal{C}_{L-1}}{x}+\frac{\mathcal{D}_{L-1}}{x^2}\right)^2\right]
        +o\!\left(\frac{1}{x^3}\right)
        \approx 
        1 - \frac{2L-1}{2x} + \frac{(2L-1)(2L-3)}{8x^2}.
    \end{aligned}
\end{equation}
\vspace*{-0.9cm}
\end{figure*}

For intermediate values of \(x\), to balance approximation accuracy and numerical stability, 
we adopt the closed-form approximation proposed in 
\cite{39.Segura2023SimpleBoundsWithBestPossibleAccuracyForRatiosOfModifiedBesselFunctions} to represent \(\mathcal{R}_L\left(x\right)\), namely,
\begin{equation}\label{eq:equ99}  
    \mathcal{R}_L\left(x\right)
    \approx
    \frac{x}{L - 0.5 + \sqrt{\left(L - 0.5\right)^2 + x^2}}.
\end{equation}

Based on the regional approximations of \(\mathcal{R}_L(x)\), a unified piecewise approximation is constructed in \eqref{eq:equ100}. 
This analytical form is then used to replace the direct numerical evaluation of \(\mathcal{R}_L(x)\), enabling fast construction of the GLRT statistic.
\begin{equation}\label{eq:equ100}  
    \begin{cases}
        \frac{x}{2L}\left[ 1 - \frac{x^2}{4L\left(L+1\right)} + \frac{x^4}{8L^2\left(L+1\right)\left(L+2\right)} \right],
        & x \to 0, \\
        1 - \frac{2L-1}{2x} + \frac{\left(2L-1\right)\left(2L-3\right)}{8x^2},
        & x \to \infty, \\
        \frac{x}{L - 0.5 + \sqrt{\left(L - 0.5\right)^2 + x^2}},
        & \text{otherwise}.
    \end{cases}
\end{equation}
\subsection{Initialization and Stability Enhancement}
To enhance the convergence and numerical stability of SA-GLRT-D, we introduce initialization strategies for the background noise mean, 
target amplitude \(\alpha\), and offset parameters \((\Delta r,\Delta v)\). 
The noise mean \(\mu_z\) is initialized using our previous method \cite{35.Zhu2025CTCFARRobustCFARDetector}, 
while \(\alpha\) is obtained from frequency refinement of the CUT spectral peak and \((\Delta r,\Delta v)\) are initialized by local full-template fitting.

Specifically, \((\Delta r,\Delta v)\) denote the range- and velocity-bin offsets between the true CUT spectral peak and its nearest discrete sampling point. 
Their initial values are defined as
\begin{equation}\label{eq:equ101}  
    \Delta r = \frac{1}{L}\sum_{l=1}^L \delta_{\mathrm{R}l},
    \quad
    \Delta v = \frac{1}{L}\sum_{l=1}^L \delta_{\mathrm{V}l},
\end{equation}
where \(\delta_{\mathrm{R}l}\) and \(\delta_{\mathrm{D}l}\) denote the range- and Doppler-bin offsets of the true CUT spectral peak from 
its nearest discrete sampling point for the \(l\)-th receive channel. 
They are estimated using the Candan frequency-refinement algorithm 
\cite{40.Candan2015FineResolutionFrequencyEstimationFromThreeDFTSamplesCaseOfWindowedData, 41.Candan2011AMethodForFineResolutionFrequencyEstimation}, 
where the offset \(\delta\) is given by
\begin{equation}\label{eq:equ102}  
    \delta = \eta \,\mathrm{Re}\!
    \left[\frac{y\left(n_k+1\right) - y\left(n_k-1\right)}
    {2y\left(n_k\right) - y\left(n_k+1\right) - y\left(n_k-1\right)}\right],
\end{equation}
where \(\eta\in\{1,2,\ldots\}\) denotes the window-dependent adjustment coefficient, 
and \(y\left(n_k\right)\) denotes the complex spectral sample at index \(n_k\). 
For each channel, three adjacent CUT samples along the fast- and slow-time dimensions are substituted into 
\eqref{eq:equ102} to obtain \(\delta_{\mathrm{R}l}\) and \(\delta_{\mathrm{V}l}\).

Since \(\alpha\) enters the model only through \(\rho=|\alpha|^2\), its initialization is reduced to that of \(\rho\). 
The initial \(\rho\) is obtained by fitting a CUT-centered local full-target NCA power-spectrum template to the observation region. 
Specifically, define
\(\bm{\mathcal{P}}^{(r_{\mathrm D},v_{\mathrm D})}\in\mathbb{R}^{(N_{\mathrm r}N_{\mathrm v})\times(N_{\mathrm r}N_{\mathrm v})}\) as
\begin{equation}\label{eq:equ103}  
    \begin{aligned}
        &\bm{\mathcal{P}}^{\left( r_{\mathrm D},v_{\mathrm D} \right)} = \\
        &\left[
            \bm{p}_1^{\left( r_{\mathrm D}-N_{\mathrm r},\,v_{\mathrm D}-N_{\mathrm v} \right)}, \ldots,
            \bm{p}_k^{\left( r_{\mathrm D},\,v_{\mathrm D} \right)}, \ldots,
            \bm{p}_{N_{\mathrm r}N_{\mathrm v}}^{\left( r_{\mathrm D}+N_{\mathrm r},\,v_{\mathrm D}+N_{\mathrm v} \right)}
        \right],
    \end{aligned}
\end{equation}
where \(\bm{p}_k^{\left( r_{\mathrm D},v_{\mathrm D} \right)}\in\mathbb{R}^{N_{\mathrm r}N_{\mathrm v}}\) denotes the NCA power-spectrum vector corresponding to the \(k\)-th target. According to \eqref{eq:equ8_templateOfCUT}, it can be expressed as
\begin{equation}\label{eq:equ104}  
    \bm{p}_k^{\left( r_{\mathrm D},v_{\mathrm D} \right)}
    =
    \operatorname{vec}\!\left(
        L\left|
            \mathscr{Y}_{0,0}^{\left( r_{\mathrm D},v_{\mathrm D} \right)}\left[\tilde p,\tilde q\right]
        \right|^2
    \right),
\end{equation}
where \(\operatorname{vec}(\cdot)\) denotes the vectorization operator.

Based on the above formulation, the optimal amplitude coefficient vector
\(\hat{\bm{\rho}}_{\mathrm A}\in\mathbb{R}^{N_{\mathrm r}N_{\mathrm v}}\)
is obtained by solving the least-squares problem and can be expressed in closed form as
\begin{equation}
    \hat{\bm{\rho}}_{\mathrm A}
    =
    \arg\min_{\bm{\rho}_{\mathrm A}}
    \left\|
        \bm{Z} - \bm{P}^{(r_{\mathrm D},v_{\mathrm D})} \bm{\rho}_{\mathrm A}
    \right\|_2^2
    =
    \tilde{\bm{P}}^{(r_{\mathrm D},v_{\mathrm D})}\bm{Z},
\end{equation}
where \(\tilde{\bm{P}}^{(r_{\mathrm D},v_{\mathrm D})}\) is the Moore--Penrose pseudoinverse of 
\(\bm{P}^{(r_{\mathrm D},v_{\mathrm D})}\).

Finally, the entry of \(\hat{\bm{\rho}}_{\mathrm A}\) corresponding to the CUT is taken as the initial value of the parameter \(\rho\).
\section{Experimental Results and Performance Evaluation}
This section evaluates the proposed SA-GLRT-D and SA-GLRT-D\(_{\mathrm{F}}\) using simulation and measured FMCW radar data. 
The metrics include \(P_{\mathrm{d}}\), \(P_{\mathrm{fa}}\), \(P_{\mathrm{a}}\), \(F_1\), 
the RMSEs of \(\rho\), \(\Delta r\), and \(\Delta v\), and computational complexity. 
We assess detection and estimation performance under additive complex Gaussian noise, 
validate practical performance using real-scene measurements, and compare asymptotic complexity and runtime among different detectors.

All metrics are averaged over 500 independent Monte Carlo (MC) trials. 
For each SNR, the noise power is set according to the average power of the noise-free echo signal. 
Each simulated echo contains 100 targets with randomly distributed ranges and velocities, with zero-velocity targets accounting for no more than half of all targets. 
The main simulation parameters are listed in Table \ref{tab1}.
\begin{table}[t]  
    \centering
    \caption{Simulation Parameters}
    \label{tab1}
    \begin{tabular}{
        >{\raggedright\arraybackslash}p{0.6\columnwidth}
        >{\raggedright\arraybackslash}p{0.25\columnwidth}}
    \toprule
    Parameter & Value \\
    \midrule
    Starting frequency & 77 \(\mathrm{GHz}\) \\
    Sweep slope & 120 \(\mathrm{MHz}/\mu\mathrm{s}\) \\
    Sweep bandwidth & 3.413 \(\mathrm{GHz}\) \\
    Sampling rate & 9000 \(\mathrm{Ksps}\) \\
    Number of samples per chirp & 256 \\
    Number of chirps per frame & 64 \\
    \bottomrule
    \end{tabular}
\end{table}
\subsection{Simulation Experiments}
\subsubsection{Detection performance versus $P_{\mathrm{FA}}$}
\begin{figure*}[!t]  
    \centering
    \includegraphics[scale=1]{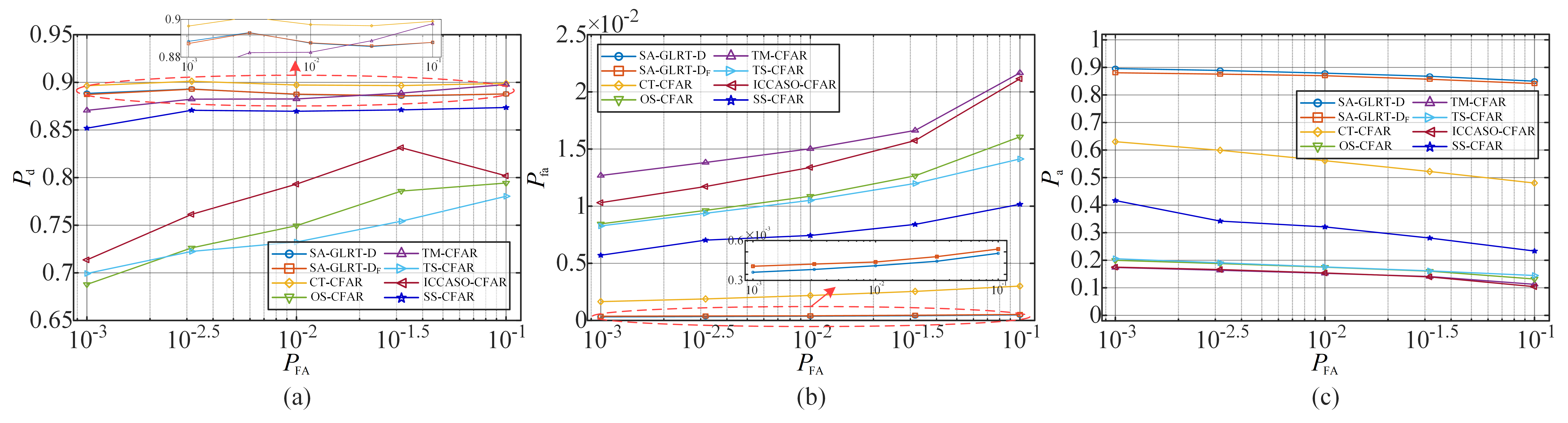}
    \caption{Performance versus \(P_{\mathrm{FA}}\) at \(\mathrm{SNR}=0\,\)dB. 
    (a) ROC curves. 
    (b) \(P_{\mathrm{fa}}\) versus \(P_{\mathrm{FA}}\). 
    (c) \(P_{\mathrm a}\) versus \(P_{\mathrm{FA}}\).}
    \label{fig6}
\end{figure*}

Fig. \ref{fig6} shows the effect of the prescribed false-alarm probability \(P_{\mathrm{FA}}\) on the performance of the SA-GLRT-D family and several advanced detectors under \(\mathrm{SNR}=0\,\mathrm{dB}\). 
Fig. \ref{fig6}(a) depicts the ROC curves of the compared detectors, where all algorithms achieve favorable detection performance around \(P_{\mathrm{FA}}=10^{-1.5}\). 
Fig. \ref{fig6}(b) and Fig. \ref{fig6}(c) depict the variations of \(P_{\mathrm{fa}}\) and \(P_{\mathrm a}\) with respect to \(P_{\mathrm{FA}}\), respectively. 
The results indicate that the proposed SA-GLRT-D family is less sensitive to \(P_{\mathrm{FA}}\) than the other detectors, demonstrating better robustness and more stable overall performance. 
Accordingly, \(P_{\mathrm{FA}}=10^{-1.5}\) is adopted in the subsequent experiments.

\subsubsection{Detection performance versus SNR under different initialization strategies}

\begin{figure*}[!t]  
    \centering
    \includegraphics[scale=1]{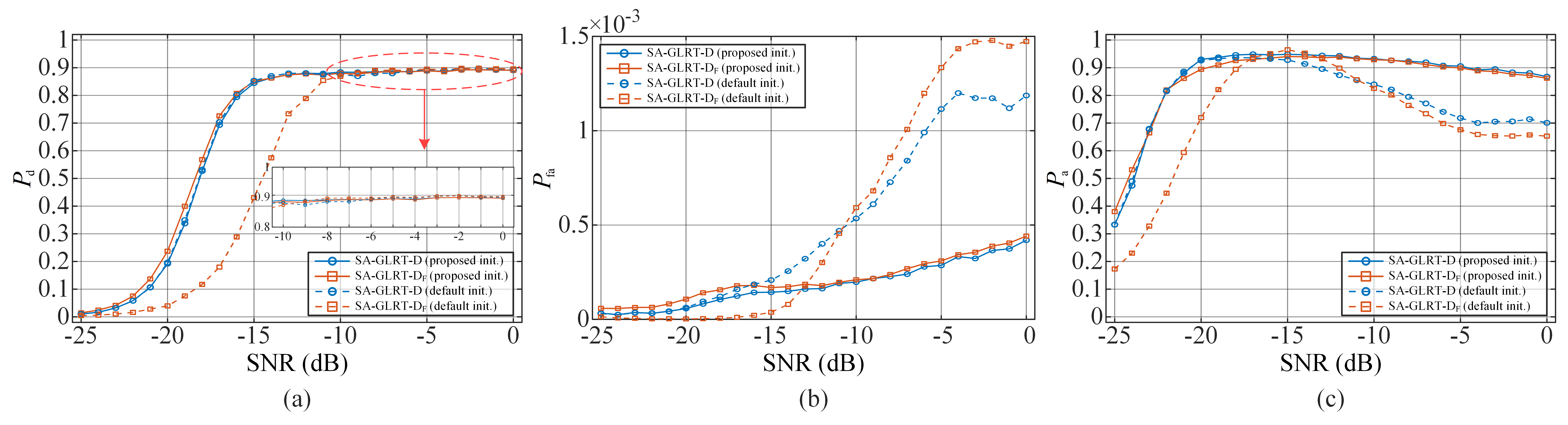}
    \caption{Performance of the SA-GLRT-D family versus SNR under different initialization strategies at \(P_{\mathrm{FA}}\!=\!10^{-1.5}\). 
    (a) \(P_{\mathrm{d}}\) versus SNR. 
    (b) \(P_{\mathrm{fa}}\) versus SNR. 
    (c) \(P_{\mathrm{a}}\) versus SNR.}
    \label{fig16}
\end{figure*}

To assess the effect of parameter initialization on the numerical stability of the SA-GLRT-D family, 
we compare the proposed initialization strategy with a direct scheme that uses the CUT index and the corresponding observed power as initial values. 
Fig. \ref{fig16} shows the performance of the SA-GLRT-D family versus SNR under the two initialization schemes at \(P_{\mathrm{FA}}=10^{-1.5}\).

Fig. \ref{fig16}(a) depicts the detection-probability results, where SA-GLRT-D is relatively insensitive to initialization, 
whereas SA-GLRT-D\(_{\mathrm{F}}\) benefits more from the proposed strategy. 
Fig. \ref{fig16}(b) and Fig. \ref{fig16}(c) depict the false-alarm and detection-accuracy results, respectively. 
Proper initialization improves false-alarm control and detection accuracy, 
whereas the direct initialization scheme degrades the overall detection performance. 
These results confirm that the proposed initialization strategy improves the numerical stability of the SA-GLRT-D family.

\subsubsection{Parameter estimation RMSEs}
\begin{figure}[!t]  
    \centering
    \includegraphics[scale=1]{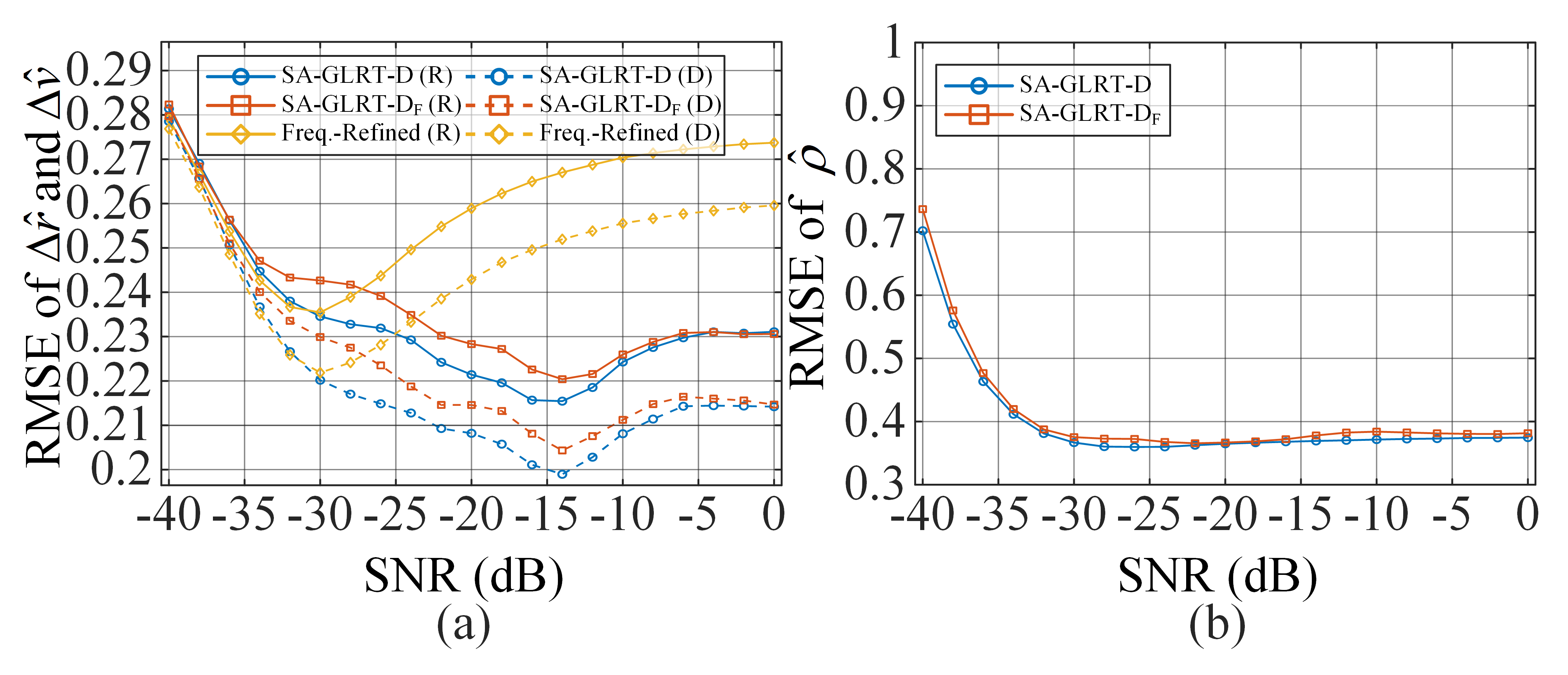}
    \caption{RMSEs of \(\Delta \hat r\), \(\Delta \hat v\), and \(\hat \rho\) versus SNR at \(P_{\mathrm{FA}}\!=\!10^{-1.5}\). 
    Dashed lines denote the overall mean RMSE across all SNR values. 
    (a) RMSE of \(\Delta \hat r\) and \(\Delta \hat v\) versus SNR.
    (b) RMSE of \(\hat \rho\) versus SNR.}
    \label{fig7}
\end{figure}

Fig. \ref{fig7} shows the RMSEs of the estimated parameters, \(\Delta \hat r\), \(\Delta \hat v\), and \(\hat \rho\), 
for the proposed SA-GLRT-D family as functions of SNR under \(P_{\mathrm{FA}} = 10^{-1.5}\). 
Fig. \ref{fig7}(a) depicts the RMSEs of \(\Delta \hat r\) and \(\Delta \hat v\). 
As the SNR increases, the estimation errors of the proposed algorithms generally decrease, indicating improved parameter estimation accuracy. 
In contrast, the errors of the frequency-refinement algorithm first decrease and then increase, showing a nonmonotonic behavior. 
For the proposed algorithms, small fluctuations can be observed in the medium-to-high SNR region. 
A possible explanation is that, under the two-stage grid-search-based estimation mechanism, the dominant error source shifts from random noise at low SNR to the bias caused by discrete grid resolution at high SNR. 
Fig. \ref{fig7}(b) depicts the RMSE of \(\hat \rho\) versus SNR. 
The results indicate that the estimation error of \(\hat \rho\) decreases with increasing SNR and converges to around \(-30\,\mathrm{dB}\). 
Similar to the errors of \(\Delta \hat r\) and \(\Delta \hat v\), slight fluctuations remain after convergence. 
Overall, Fig. \ref{fig7} demonstrates that the proposed SA-GLRT-D family can effectively estimate the CUT parameters, 
with estimation errors generally decreasing as the SNR increases.

\subsubsection{Detection performance versus SNR}
\begin{figure*}[!t]  
    \centering
    \includegraphics[scale=1]{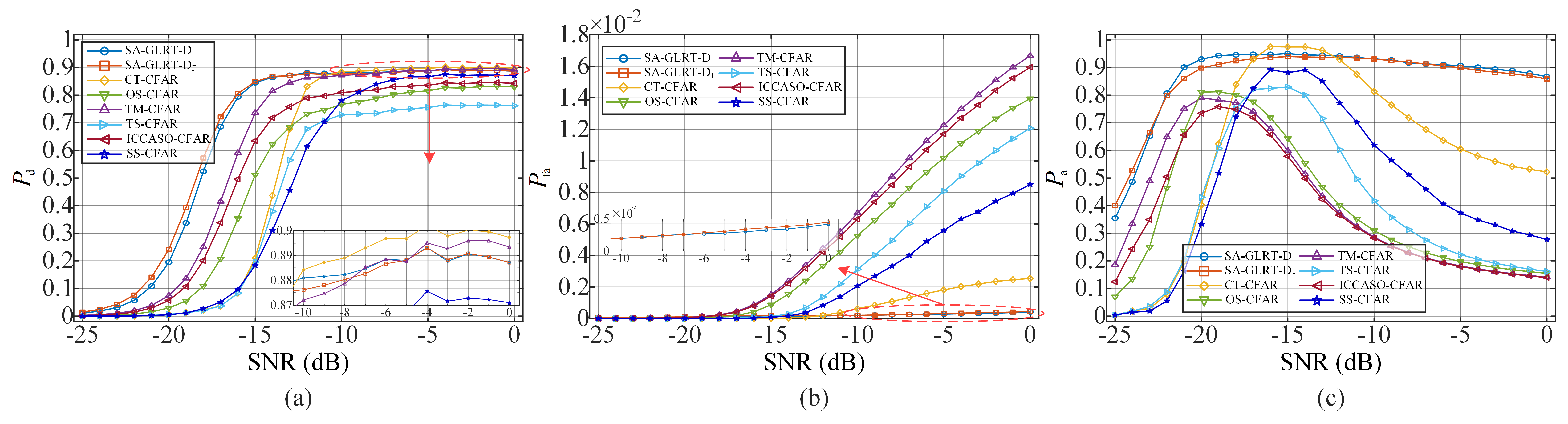}
    \caption{Performance versus SNR at \(P_{\mathrm{FA}}\!=\!10^{-1.5}\). 
    (a) \(P_{\mathrm{d}}\) versus SNR. 
    (b) \(P_{\mathrm{fa}}\) versus SNR. 
    (c) \(P_{\mathrm{a}}\) versus SNR.}
    \label{fig8}
\end{figure*}

Fig. \ref{fig8} shows the detection performance of the proposed SA-GLRT-D family and other advanced detectors versus SNR 
in a multi-target scenario under \(P_{\mathrm{FA}}=10^{-1.5}\). 
Fig. \ref{fig8}(a) depicts the detection probability \(P_{\mathrm{d}}\) versus SNR. 
In the low-SNR region \((\mathrm{SNR}<-10\,\mathrm{dB})\), the proposed SA-GLRT-D family still outperforms the other advanced detectors, despite a slight performance degradation. 
As the SNR increases, the \(P_{\mathrm{d}}\) curves of all detectors gradually converge. 
Among them, CT-CFAR achieves the highest \(P_{\mathrm{d}}\), while the performance gap between the proposed SA-GLRT-D family and CT-CFAR decreases to less than \(1\%\).
Fig. \ref{fig8}(b) shows the false-alarm probability \(P_{\mathrm{fa}}\) versus SNR. 
SA-GLRT-D achieves the lowest \(P_{\mathrm{fa}}\), and its performance remains stable as the SNR increases, demonstrating strong robustness against false alarms. 
Fig. \ref{fig8}(c) shows the probability of accurate detection \(P_{\mathrm{a}}\) versus SNR. 
As the SNR increases, the \(P_{\mathrm{a}}\) values of all detectors first increase and then gradually stabilize. 
Except for CT-CFAR at \(-15\,\mathrm{dB}\), SA-GLRT-D consistently achieves the best \(P_{\mathrm{a}}\). 
Overall, Fig. \ref{fig8} demonstrates that SA-GLRT-D provides a favorable tradeoff among detection probability, false-alarm suppression, and accurate detection performance.

Table \ref{tab2} summarizes the average detection performance of all algorithms in the SNR simulation experiments. 
All values are reported to three decimal places, and the best result for each metric is highlighted in bold. 
The results indicate that SA-GLRT-D achieves the lowest \(P_{\mathrm{fa}}\) and the highest \(P_{\mathrm{a}}\), 
demonstrating its advantages in false-alarm suppression and accurate detection. 
In contrast, SA-GLRT-D\(_\mathrm{F}\) achieves the highest \(P_{\mathrm{d}}\) and \(F_1\). 
Compared with SA-GLRT-D, SA-GLRT-D\(_\mathrm{F}\) increases \(P_{\mathrm{fa}}\) by only about \(15.8\%\), while reducing \(P_{\mathrm{a}}\) by only about \(0.7\%\). 
Considering all metrics, SA-GLRT-D\(_\mathrm{F}\) provides a more favorable overall tradeoff.
\begin{table}[t]  
    \centering
    \caption{Average Performance in SNR Simulation Experiments}
    \label{tab2}
    \begin{tabular}{lcccc}
    \toprule
    Method & $P_{\mathrm{d}}$ & $P_{\mathrm{fa}}$ $(\times 10^{-3})$ & $P_{\mathrm{a}}$ & $F_1$ \\
    \midrule
    SA-GLRT-D      & 0.648 & \textbf{0.182} & \textbf{0.868} & 0.690 \\
    SA-GLRT-D$_{\mathrm{F}}$ & \textbf{0.658} & 0.210 & 0.862 & \textbf{0.700} \\
    CT-CFAR        & 0.501 & 0.754 & 0.606 & 0.458 \\
    OS-CFAR        & 0.594 & 5.830 & 0.406 & 0.323 \\
    TM-CFAR        & 0.499 & 4.717 & 0.394 & 0.291 \\
    TS-CFAR        & 0.421 & 3.504 & 0.383 & 0.270 \\
    ICCASO-CFAR    & 0.543 & 5.542 & 0.377 & 0.302 \\
    SS-CFAR        & 0.456 & 2.401 & 0.467 & 0.349 \\
    \bottomrule
    \end{tabular}
\end{table}

\subsubsection{Detection performance versus the number of targets}
\begin{figure*}[!t]  
    \centering
    \includegraphics[scale=1]{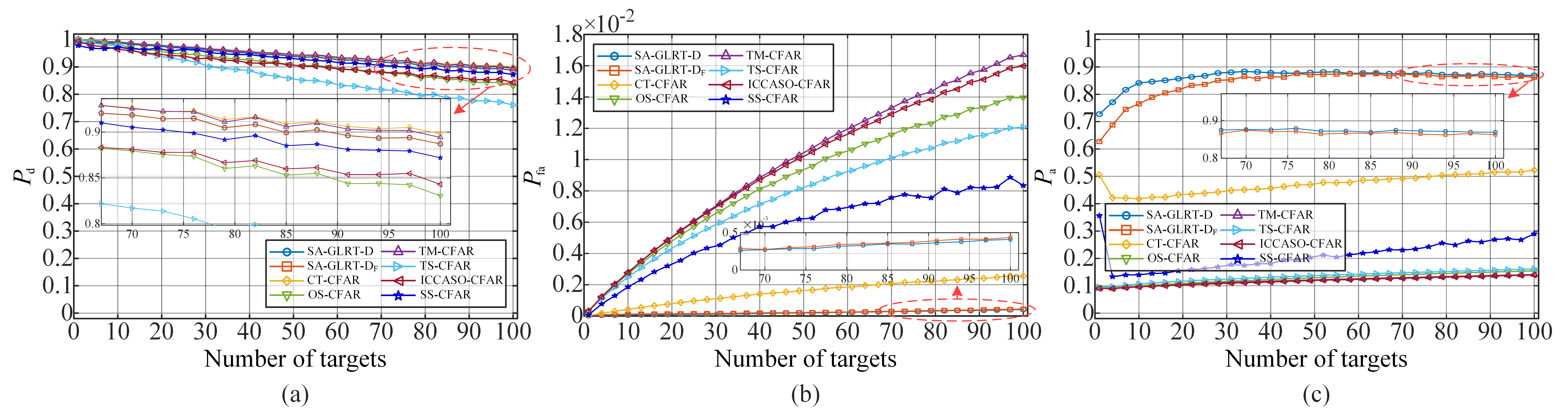}
    \caption{Performance versus the number of targets at \(P_{\mathrm{FA}}\!=\!10^{-1.5}\) and \(\mathrm{SNR}=0\,\mathrm{dB}\). 
    (a) \(P_{\mathrm{d}}\) versus number of targets. 
    (b) \(P_{\mathrm{fa}}\) versus number of targets. 
    (c) \(P_{\mathrm{a}}\) versus number of targets.}
    \label{fig9}
\end{figure*}

Fig. \ref{fig9} shows the detection performance of the SA-GLRT-D family and other advanced detectors versus the number of targets under 
\(P_{\mathrm{FA}}=10^{-1.5}\) and \(\mathrm{SNR}=0\,\mathrm{dB}\). 
Fig. \ref{fig9}(a) depicts the detection probability \(P_{\mathrm{d}}\) versus the number of targets. 
As the number of targets increases, \(P_{\mathrm{d}}\) gradually decreases for all detectors. 
CT-CFAR achieves the highest overall \(P_{\mathrm{d}}\), while the performance gap between the proposed SA-GLRT-D family and CT-CFAR is only about \(1\%\). 
Fig. \ref{fig9}(b) depicts the false-alarm probability \(P_{\mathrm{fa}}\) versus the number of targets. 
The proposed SA-GLRT-D family achieves the lowest overall \(P_{\mathrm{fa}}\), 
whereas the \(P_{\mathrm{fa}}\) values of the other detectors increase significantly as the number of targets grows. 
Fig. \ref{fig9}(c) depicts the probability of accurate detection \(P_{\mathrm{a}}\) versus the number of targets. 
The \(P_{\mathrm{a}}\) values of CT-CFAR and SS-CFAR first decrease rapidly and then increase slowly, while those of the other detectors show a generally increasing trend. 
Among them, the proposed SA-GLRT-D family exhibits the fastest improvement in \(P_{\mathrm{a}}\) and quickly stabilizes at around \(90\%\), 
which is significantly higher than those of the other advanced detectors. 

Overall, Fig. \ref{fig9} demonstrates that the proposed SA-GLRT-D family provides more stable detection performance across different target-number scenarios, 
especially in terms of false-alarm suppression and accurate detection.

\subsection{Measured Data Experiments}
\begin{figure}[!t]  
    \centering
    \includegraphics[scale=1]{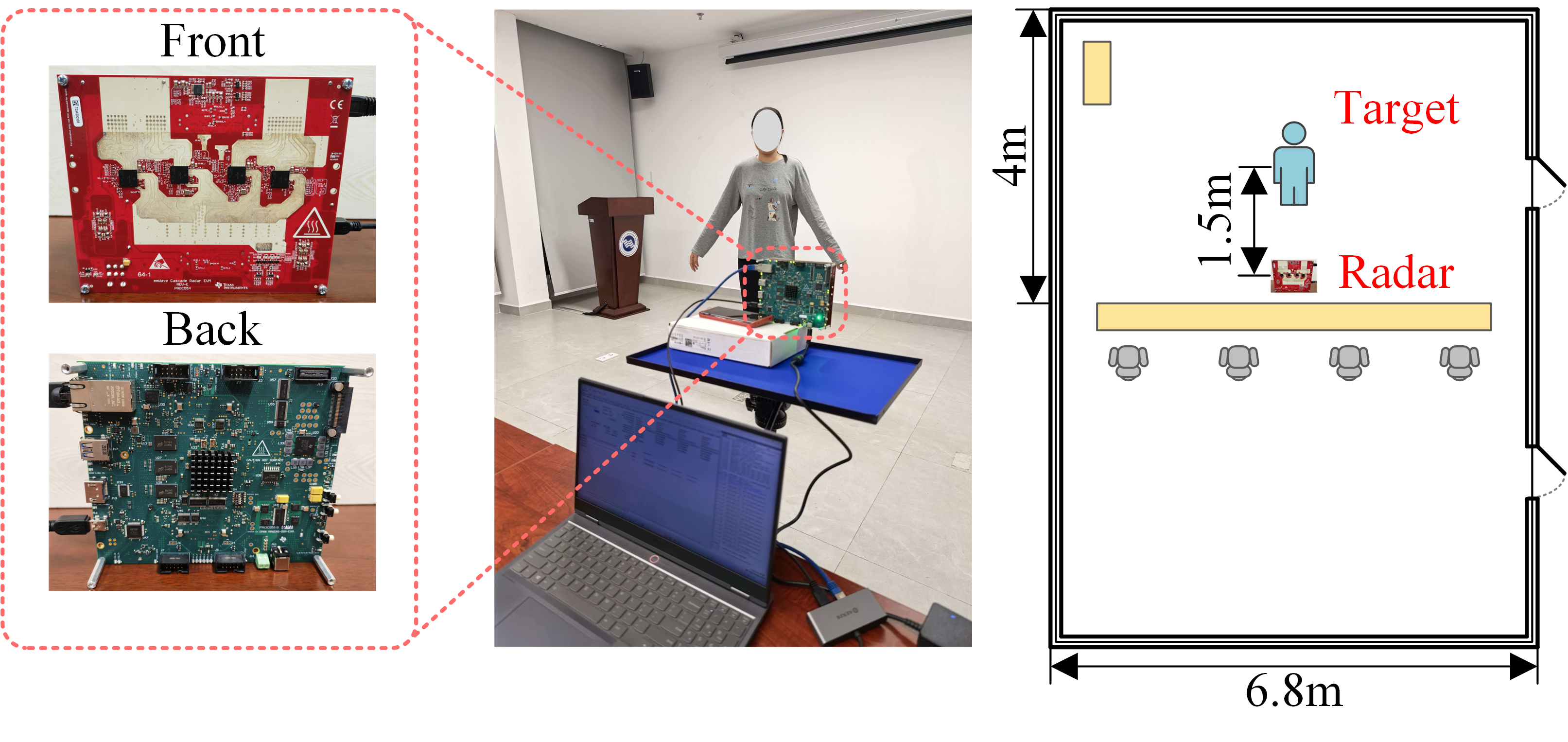}
    \caption{Indoor measurement site and radar hardware configuration. The subject stands with arms half-raised.}
    \label{fig10}
\end{figure}

A classroom experiment is conducted to evaluate the real-world detection performance of the proposed SA-GLRT-D family. 
The experimental area is \(6.8\,\mathrm{m}\times4\,\mathrm{m}\), with the subject standing \(1.5\,\mathrm{m}\) from the radar in a half-raised-arm posture. 
Data are collected using the Texas Instruments AWR2243 four-chip cascaded radar 
\cite{42.TexasInstrumentsAWRxCascadedRadarRFEvaluationModuleMMWCASRFEVM}, mounted on a tripod at a height of approximately \(1.1\,\mathrm{m}\). 
The radar parameters are listed in Table \ref{tab3}, and the acquisition scene is shown in Fig. \ref{fig10}.
\begin{table}[t]  
    \centering
    \caption{Radar Configuration Parameters}
    \label{tab3}
    \begin{tabular}{
        >{\raggedright\arraybackslash}p{0.6\columnwidth}
        >{\raggedright\arraybackslash}p{0.25\columnwidth}}
    \toprule
    Parameter & Value \\
    \midrule
    Starting frequency & 77 \(\mathrm{GHz}\) \\
    Sweep slope & 120 \(\mathrm{MHz}/\mu\mathrm{s}\) \\
    Sweep bandwidth & 3.413 \(\mathrm{GHz}\) \\
    Sampling rate & 9000 \(\mathrm{Ksps}\) \\
    Number of samples per chirp & 256 \\
    Number of chirps per frame & 64 \\
    \bottomrule
    \end{tabular}
\end{table}

\begin{figure*}[!t]  
    \centering
    \includegraphics[scale=1]{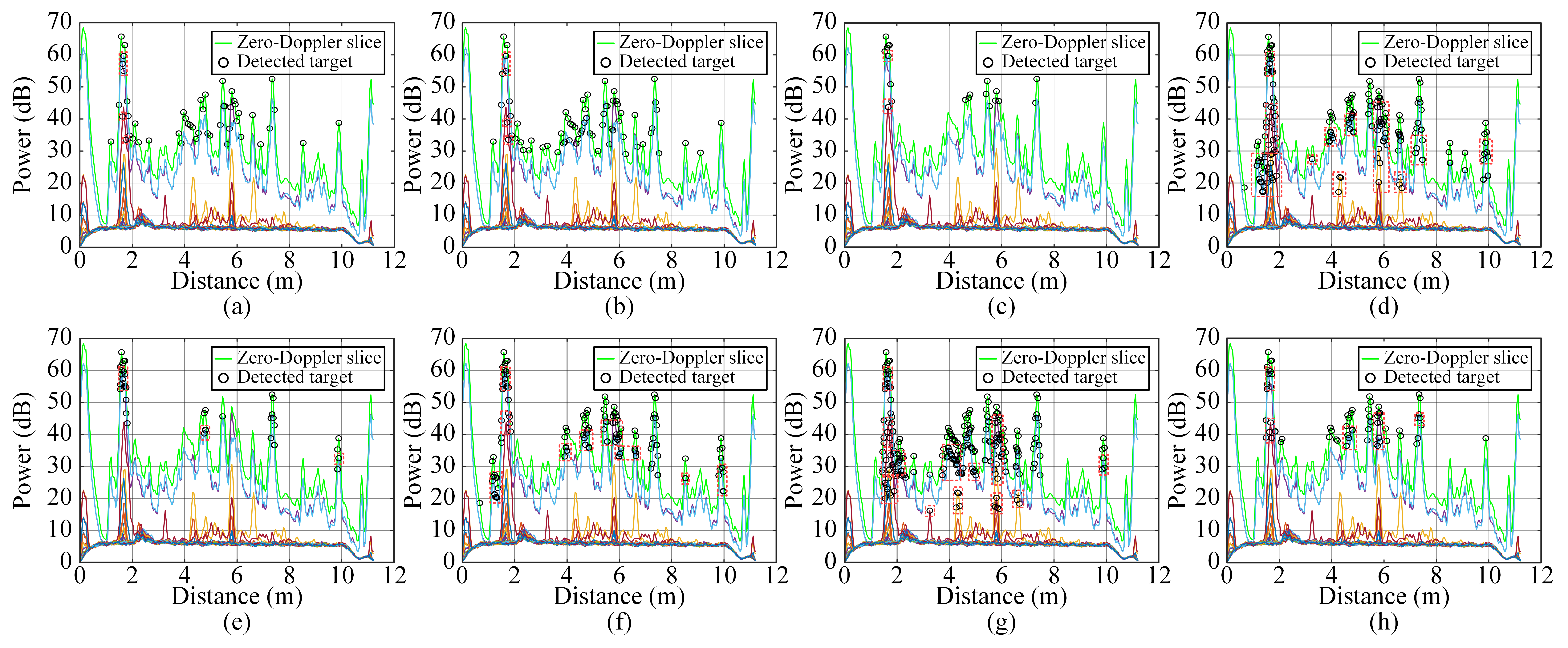}
    \caption{Range-domain side-view power responses of the NCA RDI with marked detected targets.
    (a) SA-GLRT-D. (b) SA-GLRT-D\(_\mathrm{F}\). (c) CT-CFAR. (d) OS-CFAR. (e) TM-CFAR. (f) TS-CFAR. (g) ICCASO-CFAR. (h) SS-CFAR.}
    \label{fig11}
\end{figure*}

Fig. \ref{fig11} shows the range-domain side-view power responses of the NCA RDI obtained by the proposed SA-GLRT-D family 
and the competing algorithms in the measured scenario. 
The thick green curve denotes the zero-Doppler slice, whereas the other curves correspond to Doppler slices at different velocities. 
The black circles indicate detected targets at the corresponding range--velocity cells, and the red boxes mark detections likely associated with false alarms. 
Since the measured scene contains only stationary targets, true targets are expected to lie primarily on the zero-Doppler slice.

The results show that the proposed SA-GLRT-D family, CT-CFAR, and TM-CFAR suppress false alarms effectively and achieve relatively high detection accuracy. 
For weak-target detection, the proposed SA-GLRT-D family outperforms CT-CFAR and TM-CFAR by detecting weak targets partially masked by nearby strong targets, 
such as those around \(4\,\mathrm{m}\) and \(6.5\,\mathrm{m}\). 
By contrast, OS-CFAR, TS-CFAR, and ICCASO-CFAR achieve higher detection probability at the expense of more false alarms. 
In particular, spectral-leakage responses of true targets along the velocity dimension are more likely to be falsely declared as targets, 
which degrades the overall detection performance.

\begin{figure*}[!t]  
    \centering
    \includegraphics[scale=1]{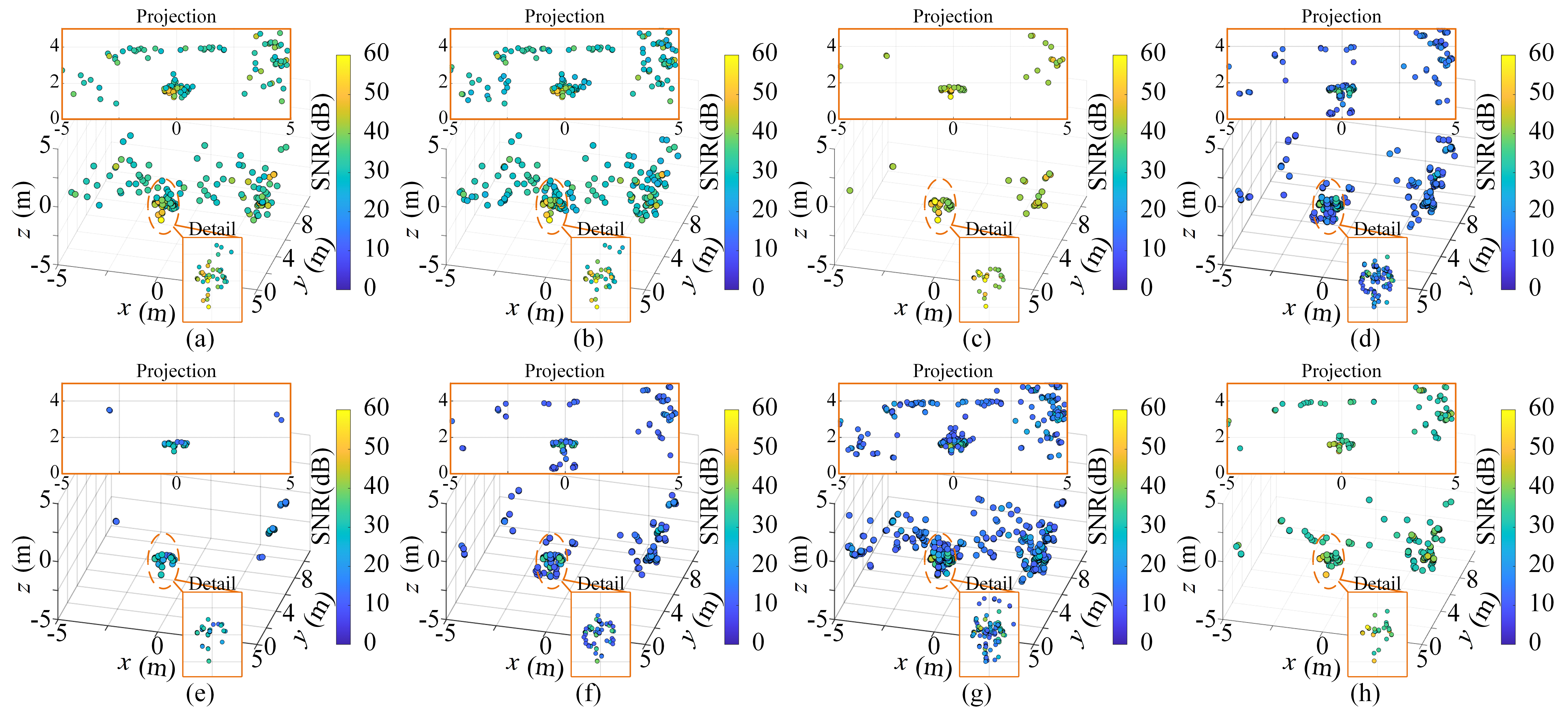}
    \caption{Generated 3D point clouds, corresponding local human-region details, and scene projections.
    (a) SA-GLRT-D. (b) SA-GLRT-D\(_\mathrm{F}\). (c) CT-CFAR. (d) OS-CFAR. (e) TM-CFAR. (f) TS-CFAR. (g) ICCASO-CFAR. (h) SS-CFAR.}
    \label{fig12}
\end{figure*}

Fig. \ref{fig12} shows the 3D point clouds generated from the detection results of the proposed SA-GLRT-D family 
and other advanced algorithms in the measured scenario using a unified angle-of-arrival estimation method. 
The boxed regions provide a detailed front view of the human-body region and an overall top view of the observed area, respectively. 
The results indicate that the proposed SA-GLRT-D family, CT-CFAR, TM-CFAR, and SS-CFAR achieve relatively high detection accuracy in the subject region, 
producing clear human-body contours, uniformly distributed point clouds, and relatively high SNRs. 
For surrounding-environment reconstruction, the SA-GLRT-D family, OS-CFAR, TS-CFAR, ICCASO-CFAR, and SS-CFAR perform better overall, 
among which SA-GLRT-D family yield clearer environmental target points and more accurate reconstruction.
\subsection{Runtime Performance Evaluation}
The computational efficiency is evaluated from theoretical and empirical perspectives. 
Table \ref{tab4:complexity_comparison} summarizes the per-iteration complexities of SA-GLRT-D, SA-GLRT-D\(_{\mathrm{F}}\), and the baseline detectors. 
The efficiency gain of SA-GLRT-D\(_{\mathrm{F}}\) mainly results from eliminating the second-stage grid search, which reduces the complexity by approximately
\begin{equation}
    \Delta \mathcal{C}
    =
    \mathcal{C}_{\mathrm{ori}}-\mathcal{C}_{\mathrm{fast}}
    \approx
    \mathcal{O}\!\left(
    K_2(N_{\mathrm r}N + N_{\mathrm v}M)
    \right).
\end{equation}
Since these expressions describe asymptotic complexity, they may not exactly match the measured runtime ranking.
\begin{table*}[t]
    \centering
    \caption{Computational Complexity Comparison of Different Detectors}
    \label{tab4:complexity_comparison}
    \resizebox{\textwidth}{!}{
    \begin{tabular}{ccc}
        \toprule
        \multicolumn{2}{c}{Item} & Computational complexity \\
        \midrule
        \multicolumn{2}{c}{CT-CFAR \cite{35.Zhu2025CTCFARRobustCFARDetector}} 
        & $\mathcal{O}\!\left(NML + NM\log(NM) + T_{\mathrm{n}}^{\mathrm{CT}}NM\log(NM)\right)$ \\
        \multicolumn{2}{c}{SS-CFAR \cite{13.Zhu2024RobustCFARDetectorBasedOnCLEANForSidelobeSuppression}} 
        & $\mathcal{O}\!\left(NM\log(NM) + T_{\mathrm{n}}^{\mathrm{SS}}NM\left[L+\log(NM)\right]\right)$ \\
        \multicolumn{2}{c}{TS-CFAR \cite{31.Ding2016RobustCFARDetectorBasedOnTruncatedStatisticsInMultipleTargetSituations}} 
        & $\mathcal{O}\!\left(NM\log(NM) + NM\right)$ \\
        \multicolumn{2}{c}{OS-Class \cite{26.Blake1988OSCFARTheoryForMultipleTargetsAndNonuniformClutter}} 
        & $\mathcal{O}\!\left(NMN_{\mathrm{ref}}\log N_{\mathrm{ref}}\right)$ \\
        \multicolumn{2}{c}{CA-Class \cite{21.Gandhi1994OptimalityOfTheCellAveragingCFARDetector}} 
        & $\mathcal{O}\!\left(NMN_{\mathrm{ref}}\right)$ \\
        \midrule
        \multirow[c]{3}{*}{\shortstack[c]{Proposed\\algorithm}}
        & per-iteration of SA-GLRT-D
        & $\mathcal{O}\!\left(NM + (N_{\mathrm p}+K_1+K_2)(N_{\mathrm r}N + N_{\mathrm v}M) + N_{\mathrm p}^3 + (I_{\hat{\gamma}} + I_{\hat{\rho}})N_{\mathrm p}\right)$ \\
        \cmidrule(lr){2-3}
        & per-iteration of SA-GLRT-D$_{\mathrm F}$
        & $\mathcal{O}\!\left(NM + (N_{\mathrm p}+K_1)(N_{\mathrm r}N + N_{\mathrm v}M) + N_{\mathrm p}^3 + (I_{\hat{\gamma}} + I_{\hat{\rho}})N_{\mathrm p}\right)$ \\
        \cmidrule(lr){2-3}
        & overall
        & $\mathcal{O}\!\left(NML + NM\log(NM) + T_{\mathrm{n}}^{\mathrm{SA}}NM\right)$ \\
        \bottomrule
    \end{tabular}}

    \vspace{2pt}
    \parbox{\textwidth}{\footnotesize\emph{Note:}
    \(N_{\mathrm p}=(2N_{\mathrm r}+1)(2N_{\mathrm v}+1)\) denotes the number of cells in the local patch.
    \(K_1\) and \(K_2\) denote the numbers of grid points used in the first- and second-stage searches, respectively.
    \(I_{\hat{\gamma}}\) and \(I_{\hat{\rho}}\) denote the numbers of iterations required to update \(\hat{\gamma}\) and \(\hat{\rho}\), respectively.
    \(T_{\mathrm n}^{\mathrm{SA}}\), \(T_{\mathrm n}^{\mathrm{SS}}\), and \(T_{\mathrm n}^{\mathrm{CT}}\) denote the iteration numbers of SA-GLRT-D, SS-CFAR, and CT-CFAR, respectively.
    \(N_{\mathrm{ref}}\) denotes the number of reference cells.}
\end{table*}

\begin{figure*}[!t]  
    \centering
    \includegraphics[scale=1]{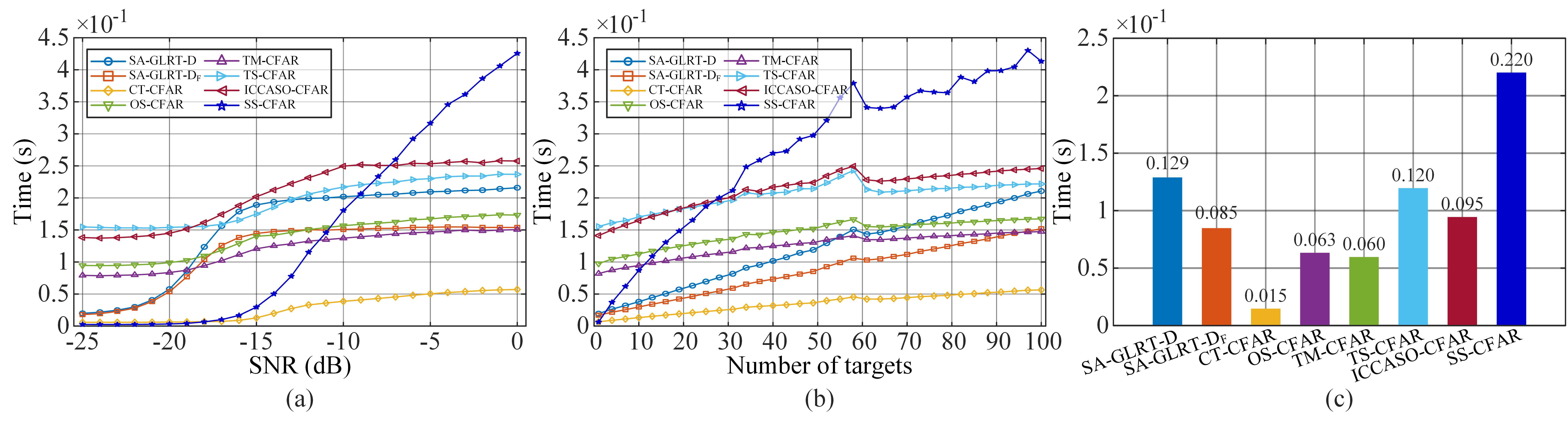}
    \caption{Runtime analysis under different experimental settings.
    (a) Runtime versus SNR in simulation experiments. 
    (b) Runtime versus the number of targets in simulation experiments. 
    (c) Runtime in the measured-data experiment.}
    \label{fig13}
\end{figure*}

To validate the above complexity analysis, we compare the runtimes of all algorithms in the SNR simulation, 
target-number simulation, and measured-data experiments, as shown in Fig. \ref{fig13}.
Fig. \ref{fig13}(a) depicts the runtime versus SNR in a multi-target scenario with 100 targets. 
CT-CFAR consistently achieves the shortest runtime, whereas SS-CFAR exhibits a rapid runtime increase as the SNR increases. 
By contrast, the runtime of the proposed SA-GLRT-D family remains relatively stable, 
with SA-GLRT-D requiring approximately \(47\%\) more runtime than SA-GLRT-D\(_{\mathrm{F}}\) on average.
Fig. \ref{fig13}(b) depicts the runtime versus the number of targets at \(\mathrm{SNR}=0\,\mathrm{dB}\). 
CT-CFAR remains the fastest method, while SA-GLRT-D\(_{\mathrm{F}}\) achieves the second-shortest runtime when the number of targets is below 100. 
SS-CFAR shows the fastest runtime growth, indicating that its computational cost is more sensitive to the target number.
Fig. \ref{fig13}(c) depicts the average runtime in the measured-data experiment, where each result is averaged over 10 independent trials. 
CT-CFAR still achieves the shortest runtime. 
Compared with CT-CFAR, SA-GLRT-D\(_{\mathrm{F}}\) improves detection accuracy with approximately twice the runtime. 
Overall, the proposed SA-GLRT-D family remains computationally acceptable and achieves a favorable tradeoff between detection performance and efficiency.

\section*{Conclusion}

This paper investigated target detection in multi-target FMCW radar scenarios with contaminated reference cells, 
CUT spectral leakage, and the resulting sidelobe interference. 
To address these problems, we proposed the SA-GLRT-D algorithm, 
which accounts for the non-ideal frequency-domain energy spreading of the CUT caused by the finite-window effect and extends conventional point-target detection to local-region detection. 
Unlike filtering-then-CFAR methods, SA-GLRT-D jointly performs decision-statistic construction, CUT-related parameter estimation, and sidelobe interference suppression within a unified hypothesis-testing framework. 
A fast implementation, SA-GLRT-D\(_{\mathrm{F}}\), is further developed to reduce computational complexity, and a parameter initialization strategy is designed to improve numerical stability. 
Experimental results showed that the proposed SA-GLRT-D family achieves high-accuracy target detection with empirically calibrated false-alarm control and favorable computational efficiency.

Future work will consider more general scenarios in which multiple scatterers coexist within the local detection region and exhibit mutual coupling and interference. 
Theoretical analysis of the parameter estimation performance, including the derivation of the corresponding Cram\'er--Rao lower bounds, is also worthy of further investigation.
\bibliographystyle{IEEEtran}
\bibliography{references}  

\end{document}